\documentclass[twocolumn]{aastex62}
\pdfoutput=1
\usepackage{amsmath,amstext,multirow}
\usepackage[T1]{fontenc}
\usepackage{apjfonts,xcolor}
\usepackage[figure,figure*]{hypcap}
\usepackage{enumitem}

\def\be{\begin{eqnarray}}   \def\ee{\end{eqnarray}}
\def\ben{\begin{equation}\begin{aligned}} \def\een{\end{aligned}\end{equation}}
\shorttitle{Velocity dispersion}
\shortauthors{Sharma et al.}
\begin{document}

\title{Fundamental relations for the velocity dispersion of stars in the Milky Way}
\author{Sanjib Sharma}
\affiliation{Sydney Institute for Astronomy, School of Physics, The University of Sydney, NSW 2006, Australia}
\affiliation{ARC Centre of Excellence for All Sky Astrophysics in Three Dimensions (ASTRO-3D)}
\author{Michael R. Hayden}
\affiliation{Sydney Institute for Astronomy, School of Physics, The University of Sydney, NSW 2006, Australia}
\affiliation{ARC Centre of Excellence for All Sky Astrophysics in Three Dimensions (ASTRO-3D)}
\author{Joss Bland-Hawthorn}
\affiliation{Sydney Institute for Astronomy, School of Physics, The University of Sydney, NSW 2006, Australia}
\affiliation{ARC Centre of Excellence for All Sky Astrophysics in Three Dimensions (ASTRO-3D)}
\author{Dennis Stello}
\affiliation{School of Physics, University of New South Wales, Sydney, NSW 2052, Australia}
\affiliation{Stellar Astrophysics Centre, Department of Physics and Astronomy, Aarhus University, DK-8000 Aarhus C, Denmark}
\affiliation{ARC Centre of Excellence for All Sky Astrophysics in Three Dimensions (ASTRO-3D)}
\author{Sven Buder}
\affiliation{Max Planck Institute  for Astronomy (MPIA), Koenigstuhl 17, D-69117 Heidelberg}
\affiliation{Research School of Astronomy \& Astrophysics, Australian National University, ACT 2611, Australia}
\affiliation{ARC Centre of Excellence for All Sky Astrophysics in Three Dimensions (ASTRO-3D)}
\author{Joel C. Zinn}
\affiliation{School of Physics, University of New South Wales, Sydney, NSW 2052, Australia}
\author{Thomas Kallinger}
\affiliation{Institute of Astrophysics, University of Vienna, Türkenschanzstrasse 17, Vienna 1180, Austria}
\author{Martin Asplund}
\affiliation{Research School of Astronomy \& Astrophysics, Australian National University, ACT 2611, Australia}
\affiliation{ARC Centre of Excellence for All Sky Astrophysics in Three Dimensions (ASTRO-3D)}
\affiliation{Fellow of the International Max Planck Research School for Astronomy \& Cosmic Physics at the University of Heidelberg}
\author{Gayandhi M. De Silva}
\affiliation{Department of Physics \& Astronomy, Macquarie University, Sydney, NSW 2109, Australia}
\author{Valentina D'Orazi}
\affiliation{INAF -Osservatorio Astronomico di Padova}
\author{Ken Freeman}
\affiliation{Research School of Astronomy \& Astrophysics, Australian National University, ACT 2611, Australia}
\author{Janez Kos}
\affiliation{Faculty of Mathematics and Physics, University of Ljubljana, Jadranska 19, 1000 Ljubljana, Slovenia}
\author{Geraint F. Lewis}
\affiliation{Sydney Institute for Astronomy, School of Physics, The University of Sydney, NSW 2006, Australia}
\author{Jane Lin}
\affiliation{Research School of Astronomy \& Astrophysics, Australian National University, ACT 2611, Australia}
\author{Karin Lind}
\affiliation{Department of Astronomy, Stockholm University, AlbaNova University Center, SE-106 91 Stockholm, Sweden}
\author{Sarah Martell}
\affiliation{School of Physics, University of New South Wales, Sydney, NSW 2052, Australia}
\affiliation{ARC Centre of Excellence for All Sky Astrophysics in Three Dimensions (ASTRO-3D)}
\author{Jeffrey D. Simpson}
\affiliation{School of Physics, University of New South Wales, Sydney, NSW 2052, Australia}
\author{Rob A. Wittenmyer}
\affiliation{Centre for Astrophysics, University of Southern Queensland, Toowoomba, Queensland 4350, Australia}
\author{Daniel B. Zucker}
\affiliation{Department of Physics \& Astronomy, Macquarie University, Sydney, NSW 2109, Australia}
\affiliation{Research Centre in Astronomy, Astrophysics \& Astrophotonics, Macquarie University, Sydney, NSW 2109, Australia}
\author{Tomaz Zwitter}
\affiliation{Faculty of Mathematics and Physics, University of Ljubljana, Jadranska 19, 1000 Ljubljana, Slovenia}
\author{Boquan Chen}
\affiliation{Sydney Institute for Astronomy, School of Physics, The University of Sydney, NSW 2006, Australia}
\author{Klemen Cotar}
\affiliation{Faculty of Mathematics and Physics, University of Ljubljana, Jadranska 19, 1000 Ljubljana, Slovenia}
\author{James Esdaile}
\affiliation{School of Physics, University of New South Wales, Sydney, NSW 2052, Australia}
\author{Marc Hon}
\affiliation{School of Physics, University of New South Wales, Sydney, NSW 2052, Australia}
\author{Jonathan Horner}
\affiliation{Centre for Astrophysics, University of Southern Queensland, Toowoomba, Queensland 4350, Australia}
\author{Daniel Huber}
\affiliation{Institute for Astronomy, University of Hawai`i, 2680 Woodlawn Drive, Honolulu, HI 96822, USA}
\author{Prajwal R. Kafle}
\affiliation{International Centre for Radio Astronomy Research (ICRAR), The University of Western Australia, 35 Stirling Highway, \\Crawley, WA 6009, Australia}
\author{Shourya Khanna}
\affiliation{Sydney Institute for Astronomy, School of Physics, The University of Sydney, NSW 2006, Australia}
\author{Yuan-Sen Ting}
\affiliation{Research School of Astronomy \& Astrophysics, Australian National University, ACT 2611, Australia}
\affiliation{Institute for Advanced Study, Princeton, NJ 08540, USA}
\affiliation{Department of Astrophysical Sciences, Princeton University, Princeton, NJ 08544, USA}
\affiliation{Observatories of the Carnegie Institution of Washington, 813 Santa Barbara Street, Pasadena, CA 91101, USA}
\author{David M. Nataf}
\affiliation{Department of Physics and Astronomy, The Johns Hopkins University, Baltimore, MD 21218, USA}
\author{Thomas Nordlander}
\affiliation{Research School of Astronomy \& Astrophysics, Australian National University, ACT 2611, Australia}
\affiliation{ARC Centre of Excellence for All Sky Astrophysics in Three Dimensions (ASTRO-3D)}
\author{Mohd Hafiz Mohd Saadon}
\affiliation{School of Physics, University of New South Wales, Sydney, NSW 2052, Australia}
\affiliation{Department of Fiqh and Usul, Academy of Islamic Studies, University of Malaya, 50603 Kuala Lumpur, Malaysia}
\author{C. G. Tinney}
\affiliation{Exoplanetary Science at UNSW, School of Physics, University of New South Wales, Sydney, NSW 2052, Australia}
\author{Gregor Traven}
\affiliation{Lund Observatory, Department of Astronomy and Theoretical Physics, Box~43, SE-221~00 Lund, Sweden}
\author{Fred Watson}
\affiliation{Department of Industry, Innovation and Science, 105 Delhi Rd, North Ryde, NSW 2113, Australia}
\author{Duncan Wright}
\affiliation{Centre for Astrophysics, University of Southern Queensland, Toowoomba, Queensland 4350, Australia}
\author{Rosemary F. G. Wyse}
\affiliation{Department of Physics and Astronomy, The Johns Hopkins University, Baltimore, MD 21218, USA}

\begin{abstract}

We explore the fundamental relations governing the radial and vertical velocity dispersions of stars in the Milky Way, from combined studies of complementary surveys including GALAH, LAMOST, APOGEE, the NASA {\it Kepler} and K2 missions, and  {\it Gaia} DR2. We find that different stellar samples, even though they target different tracer populations and employ a variety of age estimation techniques, follow the same set of fundamental relations. We provide the clearest evidence to date that, in addition to the well-known dependence on stellar age, the velocity dispersions of stars depend on orbital angular momentum $L_z$, metallicity and height above the plane $|z|$, and are well described by a multiplicatively separable functional form.
The dispersions have a power-law dependence on age with exponents of 0.441$\pm 0.007$ and 0.251$\pm 0.006$ for $\sigma_z$ and $\sigma_R$ respectively, and the power law is valid even for the oldest stars. For the solar neighborhood stars, the apparent break in the power law for older stars, as seen in previous studies, is due to the anti-correlation of $L_z$ with age. The dispersions decrease with increasing $L_z$ until we reach the Sun's orbital angular momentum, after which $\sigma_z$ increases  (implying flaring in the outer disc) while $\sigma_R$ flattens. The dispersions increase with decreasing metallicity, suggesting that the dispersions increase with birth radius. The dispersions also increase linearly with $|z|$. The same set of relations that work in the solar neighborhood also work for stars between $3<R/{\rm kpc}<20$. Finally, the high-[$\alpha$/Fe] stars follow the same relations as the low-[$\alpha$/Fe] stars.

\end{abstract}
\section{Introduction}
Most stars of the Milky Way's disc population are thought to be born on roughly circular orbits with very low velocity dispersion. However, a significant fraction of these stars are observed to have very high velocity dispersion, suggesting that they must have undergone significant dynamical evolution. Studying
the velocity distributions for disc stars can therefore not only shed light on the dynamical history of the Galaxy, but also on the dynamical processes that have shaped the present day distribution of stars. Moreover, velocity dispersion relations are an essential ingredient for constructing analytical models of the Galaxy.
Multiple studies have sought to characterize the velocity dispersion of stars in the Milky Way disc and to explain them using dynamical models. Despite much progress, however, many open questions remain.

\subsection{Background}
It has been known for a long time that the velocity dispersion of disc stars increases with age in the solar neighbourhood.
One of the earliest attempts to explain this observation dates back to \citet{1951ApJ...114..385S}, who examined in-plane
motions of disc stars.
They showed that the total dispersion of all components, $\sigma_{\rm tot}$, increases with age $\tau$ due to scattering
from massive clouds with a power-law dependence, $\tau^{\beta_{\rm tot}}$, with exponent $\beta_{\rm tot}=0.33$.
Remarkably, they inferred the presence of giant molecular clouds (GMCs) long before they were observed directly.
Later, \citet{1984MNRAS.208..687L} generalized this result by including vertical motions. However, the predictions of his model conflicted with observations. First, Lacey concluded that $\beta_{\rm tot}=0.25$,
whereas, observations suggest that solar-neighborhood stars have $\beta_{\rm tot}$ between 0.3 and 0.5 -- or more precisely that $\beta_z$ ranges from 0.35 to 0.6 and $\beta_R$ ranges from from 0.19 to 0.35
\citep{2004A&A...418..989N,2009MNRAS.397.1286A,2014ApJ...793...51S, 2019MNRAS.489..176M}.
Secondly, the overall heating rate derived (which determines the dispersion of older stars) was too low.
Finally, Lacey predicted the ratio of vertical to radial dispersion,  $\sigma_z/\sigma_R$,  to be $0.8$, which is higher than
observed ratio for stars in the solar neighborhood (0.5-0.6).

\citet{2002MNRAS.337..731H} used N-body simulations to confirm
some of the findings of \citet{1984MNRAS.208..687L}.
They found that
for scattering from GMCs, $\beta_{\rm tot}$ is indeed less than 0.33,
or more precisely that $\beta_R=0.20$, $\beta_z=0.26$ and $\beta_{\rm tot}=0.21$, which is even less than the value predicted by Lacey. The overall heating rate was also confirmed to be low.
Specifically, with the surface density of GMCs set to the
present-day value in the solar neighborhood ($5\; {\rm M_{\odot}/pc^2}$), the predicted velocity dispersion
for the oldest stars was less than that of the observations by about 15 km/s.
Additionally, \citet{2002MNRAS.337..731H} found that the ratio $\sigma_z/\sigma_R$ depends on the number density of GMCs.
For the GMC number density of $5\; {\rm M_{\odot}/pc^2}$, $\sigma_z/\sigma_R$ was $0.5$ in rough agreement with the observed value, but much less than the value of \citet{1984MNRAS.208..687L}. The value of $\sigma_z/\sigma_R$ computed by \citet{1984MNRAS.208..687L} was large
because an isotropic distribution of star-cloud impact parameters was assumed. When the anisotropy
in the impact parameters is taken into account, $\sigma_z/\sigma_R$ is 0.62 in the steady state \citep{1993MNRAS.263..875I,1999MNRAS.307..737S,2008ASPC..396..341S}.
This value of the ratio agrees well with the observations. For example, for 10 Gyr old stars,
\citet{2009MNRAS.397.1286A} report the ratio  to be 0.56  using the Geneva Copenhagen survey \citep[GCS - ][]{2004A&A...418..989N}, while \citet{2014ApJ...793...51S} report values of 0.59 and 0.65 using the GCS and RAVE surveys respectively.

The inability of GMC scattering models to match the observed data prompted
the exploration of other mechanisms to excite random motions. Transient spiral structures are one such mechanism.
They lead to potential
fluctuations in the disc that can heat up disc stars \citep{1967ApJ...150..461B,1984ApJ...282...61S, 1985ApJ...292...79C}.
\citet{2004MNRAS.350..627D} showed using numerical experiments in two dimensions that spiral arms alone can lead to $\beta_{R}$ in the range 0.25 to 0.5, and a heating rate such that the value of
$\sigma_R$ is consistent with observations.
However, spiral scattering is
too inefficient to increase the vertical dispersion
\citep{2013ApJ...769L..24S,2015ApJ...802..109M}.
This led \citet{1990MNRAS.245..305J} to argue  that a combination of spiral structure
and GMC  heating could explain the velocity dispersion of observed stars, though their predicted $\beta_z$ (0.3) was too low and the predicted $\beta_R$ (0.5) was too large.

One way to resolve the discrepancy between the predicted and the observed values of $\beta_z$ is to accommodate a scattering environment {\it that is evolving with time}. This consequently means that the velocity dispersion as a function
of age (age-velocity relation or AVR) for stars in the solar neighborhood is not same as the
evolution of velocity dispersion with time of stars born together (heating history)
at a given time in the past.
In other words, the AVR is the compilation of the end of the heating history of stellar populations born at different times.
Specifically, due to the much higher gas fraction in the early Galactic disc, the contribution of GMC scattering is expected to decrease with time, which has been shown to lead to $\beta_z$ being close to 0.25 for the heating history but greater than 0.4 for the AVR \citep{2016MNRAS.462.1697A, 2019ApJ...878...21T}.

There are other physical processes that can heat up the disc, and so shape the AVR.
The Milky Way hosts a bar
that can heat disc stars, as demonstrated in isolated disc/bar/bulge simulations \citep[e.g., ][]{2010ApJ...721.1878S}, particularly near the strong Lindblad resonances. The same effect is seen in cosmological N-body simulations, where bars emerge within the evolving disc
\citep{2016MNRAS.459..199G}. But here there is an added contribution from the disc being bombarded by orbiting satellites \citep{1999MNRAS.304..254V}. The AVR can also be shaped by the fact that the intrinsic velocity dispersion was higher at earlier times, as reported by H$\alpha$ emission of gas in external galaxies at high redshift \citep{2009ApJ...706.1364F,2015ApJ...799..209W}. However,
H$\alpha$ emission tracks ionized gas and it is not yet clear, if stars, which form out of cold gas, also have high velocity dispersion like the ionized gas.
There is now strong evidence, both theoretical \citep{2002MNRAS.336..785S,2008ApJ...684L..79R}
and observational,
that stars migrate from their place of birth.
The observational evidence comes from
the presence of low-eccentricity and super-metallicity stars in the solar neighborhood-- a realization that dates back to at least \citet{1972ade..coll...55G}\, \citep[see also][]{2015MNRAS.447.3526K,2020MNRAS.493.2952H}.
Because the heating rate is higher in
the inner regions than in the outer regions, one has to take migration into account when modelling the AVR at a given location. However, in a comprehensive review of stellar migration,  \citet{2016AN....337..703M} argues that migration, on average, generally does not lead to disc heating.

\subsection{Modelling the Physical Mechanisms of Disc Evolution}
Given that various physical processes can play a role in disc evolution, it is imperative
to study them both individually (to assess their relative importance) and in combination (to see the
full effect of them acting together). The seminal work by
\citet{2016MNRAS.462.1697A} highlights the utility of this approach. They analyzed N-body simulations that had spiral arms, GMCs,
a bar, and growing discs -- and were successful
in reproducing the AVR in the solar neighborhood of the Milky Way.
However, this model was only compared
with observations in the solar neighborhood,
and possible dependencies on metallicity and angular momentum
were not considered.
We show for the first time that
these properties taken together
can place even stronger constraints on the models.

There are also other good reasons for studying the dependence of velocity dispersion on metallicity and angular momentum.
For a given age, the metallicity provides a way to tag the birth radius of a star, which provides leverage on
the process of radial migration
\citep{2010ApJ...713..166B,2019ApJ...884...99F}.
The angular momentum $L_z$ provides the mean radius where a star spends most of its history, and so (unlike the present radius $R$) is a more useful indicator of the amount of scattering the star has undergone.
Moreover, there are strong theoretical reasons to prefer $L_z$ over $R$.
For an axisymmetric system $L_z$ is a constant of motion and, Jeans' theorem tells us that the phase space density of a system in dynamical equilibrium should only depend on constants of motion \citep{2008gady.book.....B}.
Furthermore, specifying the dispersion as a function of $L_z$ paves the way for constructing better analytical models of the Galaxy, e.g., dynamical models based on actions by \citet{2012MNRAS.426.1328B} and  \citet{2015MNRAS.449.3479S}.

Given that the velocity dispersion of a population of stars can depend
on a number of stellar properties,
it is important to come up with a useful way to characterize the velocity dispersion from observations such that it can
test theoretical models.
The selection function of a survey will tend to leave its imprint on the
measured velocity dispersions \citep{2014ApJ...793...51S}, which makes it difficult to compare and combine results from different surveys, and also to compare observed dispersions with model predictions.
If the velocity dispersion $\sigma$ only depends on a set of observables $X$, then
knowing $X$ is sufficient to characterize the dispersion
irrespective of the selection function.
This brings us to the question of identifying the fundamental relations governing the dispersion, i.e., what is a suitable choice for the
set of observables/variables $X$, and how does the velocity dispersion depend on them. Intuitively (as discussed above) the
dispersion should be governed by age, metallicity and angular momentum. {\it However, the joint dependence of dispersion on these properties has not been studied before, and this is what we address in this paper.}
Since stars migrate from their place of birth, strictly speaking,
the amount of scattering a star experiences will also depend upon the evolutionary history of its angular momentum $L_z$, but we do not consider this as we do not have any
observable that tracks this information.
Additionally, each observational survey and age estimation technique has its own systematics, and no attempt has been made to characterize such systematics, and we also address this issue.

\subsection{Disc Evolution in the Age of Massive Galactic Surveys}\label{sec:disc_evol}
A number of large observational surveys cataloging the detailed properties of a huge numbers of stars in the Milky Way mean that
we are better poised now than ever before to unravel the fundamental velocity dispersion relations. These data sets
probe  stellar kinematics well beyond the solar neighborhood, and hence provide more coverage of the angular momentum and metallicity dimensions.
Additionally, these data sets have large sample of stars that allows the additional dependence on metallicity and angular momentum to be studied robustly.
The combination of {\it Gaia} DR2 astrometry and  accurate radial velocities from ground-based spectroscopic surveys, provides precise six-dimensional phase space information for a large number of stars. Spectroscopic surveys such as GALAH and LAMOST, mean that it is now possible to get reliable age estimates for a large number of main sequence turn-off (MSTO) and subgiant stars in the solar neighborhood, a significant improvement when compared with photometry-based ages \citep[e.g.,][]{2019MNRAS.486.1167B}.
Asteroseismology from missions like {\it Kepler} and {\it K2} has opened the door to estimating the ages of intrinsically bright giant stars allowing us to study the velocity dispersions well beyond the solar neighborhood. Ground-based spectroscopic surveys also provide elemental abundances, with which we can tag stellar populations that were born at the same time and same place \citep{2002ARA&A..40..487F}.

\begin{table*}
\caption{Description of the different data sets used to study velocity dipsersion}
\begin{tabular}{lllllll}
\hline
Name & Spectroscopic survey & Stellar type & Asteroseismology  & Age estimation & Stars \\
\hline
LAMOST-MSTO & LAMOST-DR4 & MSTO &  & \citet{2017ApJS..232....2X} &  398,173 \\
LAMOST-RG-CN & LAMOST-DR4 &  red-giant-branch &  & \citet{2019MNRAS.484.5315W} & 326,606 \\
GALAH-MSTO & GALAH-iDR3 & MSTO &  & BSTEP \citep{2018MNRAS.473.2004S}& 101,328 \\
GALAH-RG-K2 & GALAH-iDR3 & red-giant & K2-CAN & BSTEP \citep{2018MNRAS.473.2004S}& 6,445\\
APOGEE-RG-KEPLER & APOGEE-DR14 & red-giant & {\it Kepler}-CAN & BSTEP \citep{2018MNRAS.473.2004S}& 6,091\\
\hline
\hline
\end{tabular}
\label{tab:datasets}
\end{table*}

Some attempts have already been made to characterize the velocity dispersions
using the large observational surveys
that probe the velocity dispersion beyond the solar neighborhood and below we summarize four such studies.
\citet{2018MNRAS.481.4093S} used data from multiple observational surveys and multiple stellar types (about 1.2 million stars), but did not consider potential systematics between the different surveys used. They studied the dependence of velocity dispersion on age and radius $R$, but ignored the dependence on angular momentum, metallicity, and $z$. They found that velocity dispersion decreases exponentially with $R$ out to the solar Galactic radius, and that beyond this  $\sigma_z$ tends to increase, while $\sigma_R$ tends to flatten out.
The velocity dispersion was found to grow as a power law with age, with exponent
$\beta_R \sim 0.3$ and $\beta_z \sim 0.4$.

\citet{2019ApJ...878...21T} and \citet{2019MNRAS.489..176M} both studied the relationship between age and kinematics
using APOGEE-DR14 data and ages estimated using a  neural network model, which was trained on asteroseismic ages from the NASA {\it Kepler} mission.
\citet{2019ApJ...878...21T} used a sample of about 20000 red clump stars
, while \citet{2019MNRAS.489..176M} used a sample of about 65000 giants.
\citet{2019ApJ...878...21T} studied the dependence
of the expectation value of the vertical action $\hat{J_z}$ as a function of age and average radius  $\overline{R}_{\rm GC}$,
which was defined to be the mean of the birth radius and the current radius. Crucially, they did not consider the dependence on birth radius and angular momentum separately, and also they did not study the in-plane kinematics.
They found that the expectation value of birth action $\hat{J}_{z,0}$ is constant until about $\overline{R}_{\rm GC}=10$ kpc
but rises beyond that.
Assuming the following approximate relations, which are valid under epicycle approximation, $\hat{J}_{z} \propto \sigma_z^2/\nu$, and $\nu \propto \sqrt{\Sigma}$, where $\nu$ is vertical oscillation frequency and $\Sigma$ is mass surface density, we interpret their results as follows.
The vertical dispersion falls off exponentially with $\overline{R}_{GC}$ until 10 kpc, but beyond that it flattens or increases.
They also found that the vertical dispersion increases with age as a power law with exponent $\beta_z$ ranging from 0.5 to 0.65.

\citet{2019MNRAS.489..176M} studied the velocity dispersion as a function of Galactocentric cylindrical coordinates $R$ and $z$ for stars binned by age, [Fe/H] and [$\alpha$/Fe].
A quadratic model was assumed for dependence on $z$, and an exponential model for dependence on $R$.
We note that the exponential model might be inappropriate,
given that \cite{2018MNRAS.481.4093S} find the dependence
of dispersion on $R$ to be exponential only for $R<8$ kpc, but flat and even rising for $R>8$ kpc.
\citet{2019MNRAS.489..176M} found that for young stars the dispersions and the ratio $\sigma_z/\sigma_R$ increase with height $|z|$, which they attribute to
stronger heating by spiral structure in the plane
and the relatively longer time scale for GMCs to
redirect random in-plane motion to vertical motion.
For a given age they find that the dispersions are
higher for those mono-metallicity populations that have
a larger mean orbital radius. But given that
they study their stars by binning them up
in mono-metallicity populations and the fact
that metallicity is anti-correlated with
mean orbital radius, we note that the observed trend with mean
orbital radius is indistinguishable from a trend with
metallicity.

\citet{2018MNRAS.481.1645M} studied the
vertical dispersion $\sigma_z$, of about 500 solar-neighborhood stars as a function of
birth radius and age. The dependence of $\sigma_z$
on angular momentum was not studied.
For a given age,
the $\sigma_z$ was found to vary with birth radius
such that it has a slope, which is positive
for old stars (age greater than 8 Gyr), flat for
intermediate age stars, and slightly negative
for young stars (age less than 4 Gyr).

\begin{table*}
\caption{Maximum likelihood estimates of parameters $\theta_v$ used to model the dispersion of velocity components $v_z$ and $v_R$. }
\begin{tabular}{lllllll}
\hline
$v$ & $\sigma_{0,v}$ & $\beta_{v}$ & $\lambda_{L,v}$ & $\alpha_{L,v}$ & $\gamma_{\rm [Fe/H],v}$ & $\gamma_{z,v}$ \\
\hline
$v_z$ & $21.1 \pm 0.2$ km/s & $0.441 \pm 0.007$ & $1130 \pm 40$ kpc km/s & $0.58 \pm 0.04$ & $-0.52 \pm 0.01$ km/s/dex & $0.20 \pm 0.01$ km/s/kpc\\
$v_R$ & $39.4 \pm 0.3$ km/s & $0.251 \pm 0.006$ & $2300 \pm 200$ kpc km/s & $0.09 \pm 0.04$ & $-0.19 \pm 0.01$ km/s/dex & $0.12 \pm 0.01$ km/s/kpc\\
\hline
\end{tabular}
\label{tab:coeff}
\end{table*}

\section{Data}
In this paper, we mainly make use of data from the LAMOST \citep{2012RAA....12..735D,2012RAA....12..723Z}
and GALAH spectroscopic survey \citep{2015MNRAS.449.2604D}.
We also use the APOGEE-DR14 spectroscopic survey
\citep{2017AJ....154...94M}, but only for the purpose
of studying systematic effects.
We used the LAMOST-DR4 value added catalog from
\citet{2017MNRAS.467.1890X}, for  radial velocity, $T_{\rm eff}$, $\log g$, [Fe/H], [$\alpha$/Fe], and distance.
For LAMOST stars, we used two types of stars, the MSTO stars
and the red-giant (RG) stars. The ages for the
LAMOST-MSTO sample were taken from \citet{2017ApJS..232....2X}
and for the LAMOST-RG-CN sample were taken from \citet{2019MNRAS.484.5315W}.
The LAMOST-RG-CN
sample consists only of red giant branch stars (red clump stars are not included),
with ages derived from spectroscopic C and N features.
For the GALAH survey, we also used two types of stars, the MSTO stars and the RG stars. More precisely, we make use of the extended GALAH catalog (GALAH+), which also includes data from TESS-HERMES \citep{2018MNRAS.473.2004S} and K2-HERMES \citep{2019MNRAS.490.5335S} surveys
that use the same spectrograph and observational setup as the GALAH survey. The RG stars that we use have asteroseismic information from the NASA K2 mission and their
spectroscopic followup was done by the K2-HERMES survey, hereafter they are referred to as GALAH-RG-K2.
We note that the spectroscopic analysis of GALAH-DR2 \citep{2018MNRAS.478.4513B} was based on a machine learning model trained on a set of 10,605 stars analysed in detail using the SME code \citep{1996A&AS..118..595V,2017A&A...597A..16P}. In this paper, we exploit parameters from GALAH-iDR3, an internal data release where every star has been analysed using SME and incorporates Gaia-DR2 distance information \citep{2018A&A...616A...1G,2018A&A...616A...2L}. A full discussion will be presented in a forthcoming paper and the results will be available as part of GALAH-DR3.

To select stars with reliable ages, we adopt the following selection function for MSTO stars,
\be
(3.2 < \log g < 4.1) \& (5000 <\; T_{\rm eff}/{\rm K}\;<7000).
\ee
For the RG stars, we adopt the following selection criteria,
\be
(1<\log g < 3.5) \& (3500\; <\; T_{\rm eff}/{\rm K}\;<5500).
\ee

The ages and distances for the GALAH-MSTO and GALAH-RG-K2
stars are computed with the BSTEP code \citep{2018MNRAS.473.2004S}.
BSTEP provides a Bayesian estimate of intrinsic stellar parameters
from observed parameters by making use of stellar isochrones.
For results presented in this paper, we use the PARSEC-COLIBRI stellar isochrones \citep{2017ApJ...835...77M}.
For the GALAH-MSTO stars, we use the following observables,
$T_{\rm eff}, \log g, [{\rm Fe/H}], [\alpha/{\rm Fe}]$, $J$, $Ks$ and parallax. For the GALAH-RG-K2 stars, in addition to the above observables, we use the asteroseismic observables $\Delta{\nu}$ and $\nu_{\rm max}$. These stars were observed by the NASA K2 mission as part of the K2GAP program \citep{2015ApJ...809L...3S} and includes stars from campaigns 1 to 15. The asteroseismic analysis is conducted with the method by \citet{2010A&A...522A...1K,2014A&A...570A..41K}, known as the CAN pipeline. $\Delta{\nu}$ and $\nu_{\rm max}$ for the model
stars in the isochrones are determined with the ASFGRID code \citet{2016ApJ...822...15S} that incorporates corrections to the
$\Delta{\nu}$ scaling relation suggested by stellar models.
A summary of different data sets used in this paper is given
in \autoref{tab:datasets}.

Transformation from heliocentric to Galactocentric coordinates is done assuming $R_{\odot}=8.0$ kpc \citep{1993ARA&A..31..345R}, $z_{\odot}=0.025$ kpc, $\Omega_{\odot}=30.24$ kms/s/kpc \citep{2004ApJ...616..872R}, $U_{\odot}=10.96$ km/s and
$W_{\odot}=7.53$ km/s \citep{2014ApJ...793...51S}.
We use a right handed $UVW$ coordinate system, where
$U$ points towards the Galactic center, $V$ is in the direction of the Galactic roation and $W$ points towards the North Galactic pole.
The transformation is carried out with
the following heliocentric quantities: Gaia-DR2 angular position and proper motions, spectroscopic radial velocities, and  spectro-photometric distances. Where needed we assume the circular velocity at Sun, $\Theta_{\odot}$, to be 232 km/s \citep{2014ApJ...793...51S}.

\begin{figure*}[tb]
\centering \includegraphics[width=0.99\textwidth]{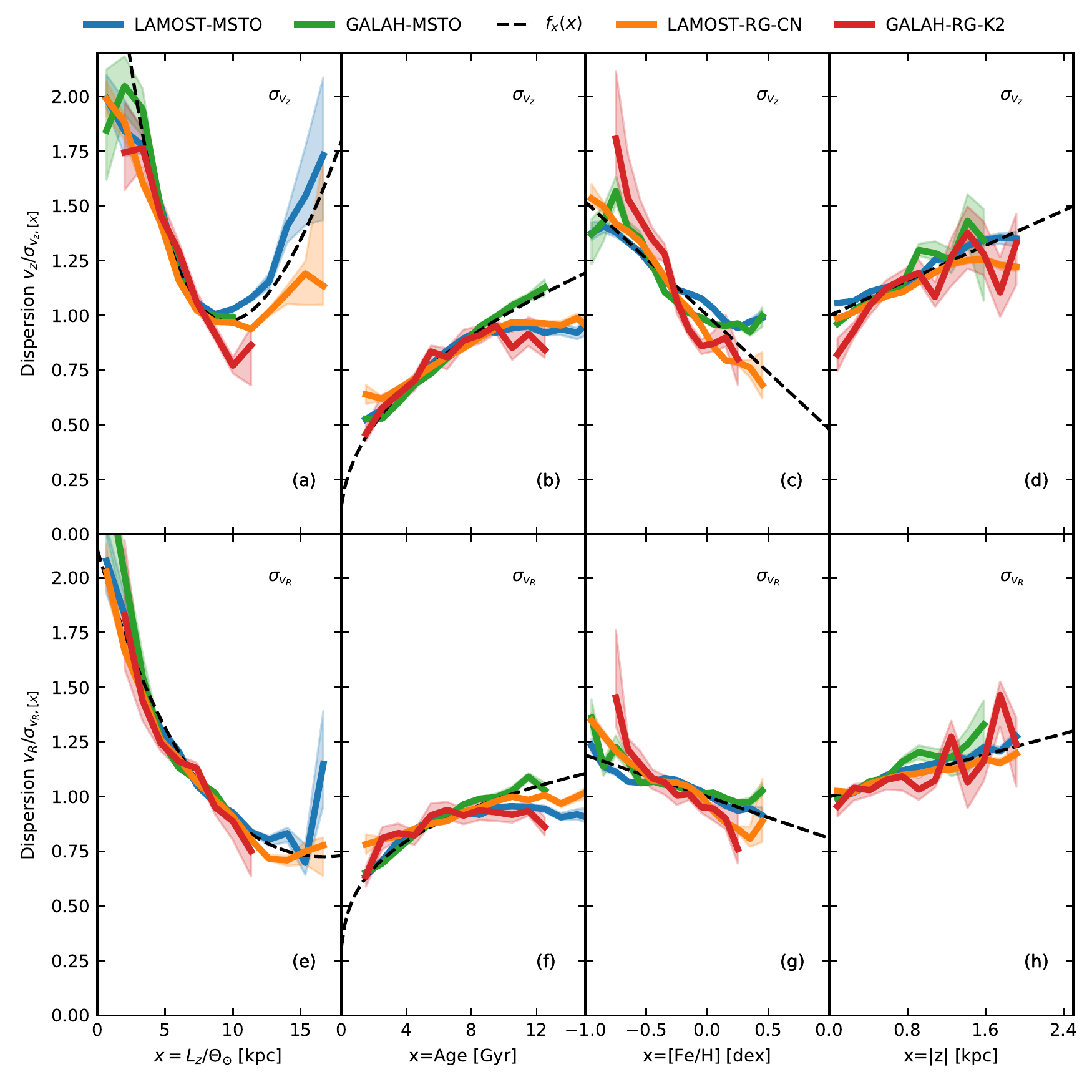}
\caption{
Velocity dispersion as a function of angular momentum, age, metallicity, and height above the Galactic midplane for different data sets. The shaded region denotes the 16 and 84 percentile confidence interval estimated using bootstrapping.
The velocity dispersion is modelled as $\sigma=\sigma_0 f_{\tau}f_{L_z}f_{{\rm [Fe/H]}} f_{z}$. Panels from left to right show the dispersion of  $v/\sigma_{v,[L_z]}$, $v/\sigma_{v,[\tau]}$, $v/\sigma_{v,\rm [Fe/H]}$, and $v/\sigma_{v,[z]}$ respectively, for the observed stars (see \autoref{equ:sigma_bracket}).
In each panel, variation with respect to other independent variables has been factored out.
The top panels show the dispersion in vertical velocity $v_z$, while the bottom panels show the dispersion in Galactocentric radial velocity $v_R$.
The dashed lines show the the best fit model profiles based on \autoref{equ:vdisp_model} and parameters given in \autoref{tab:coeff}.
\label{fig:sigma_vz_vr}}
\end{figure*}

\begin{figure}[tb]
\centering \includegraphics[width=0.49\textwidth]{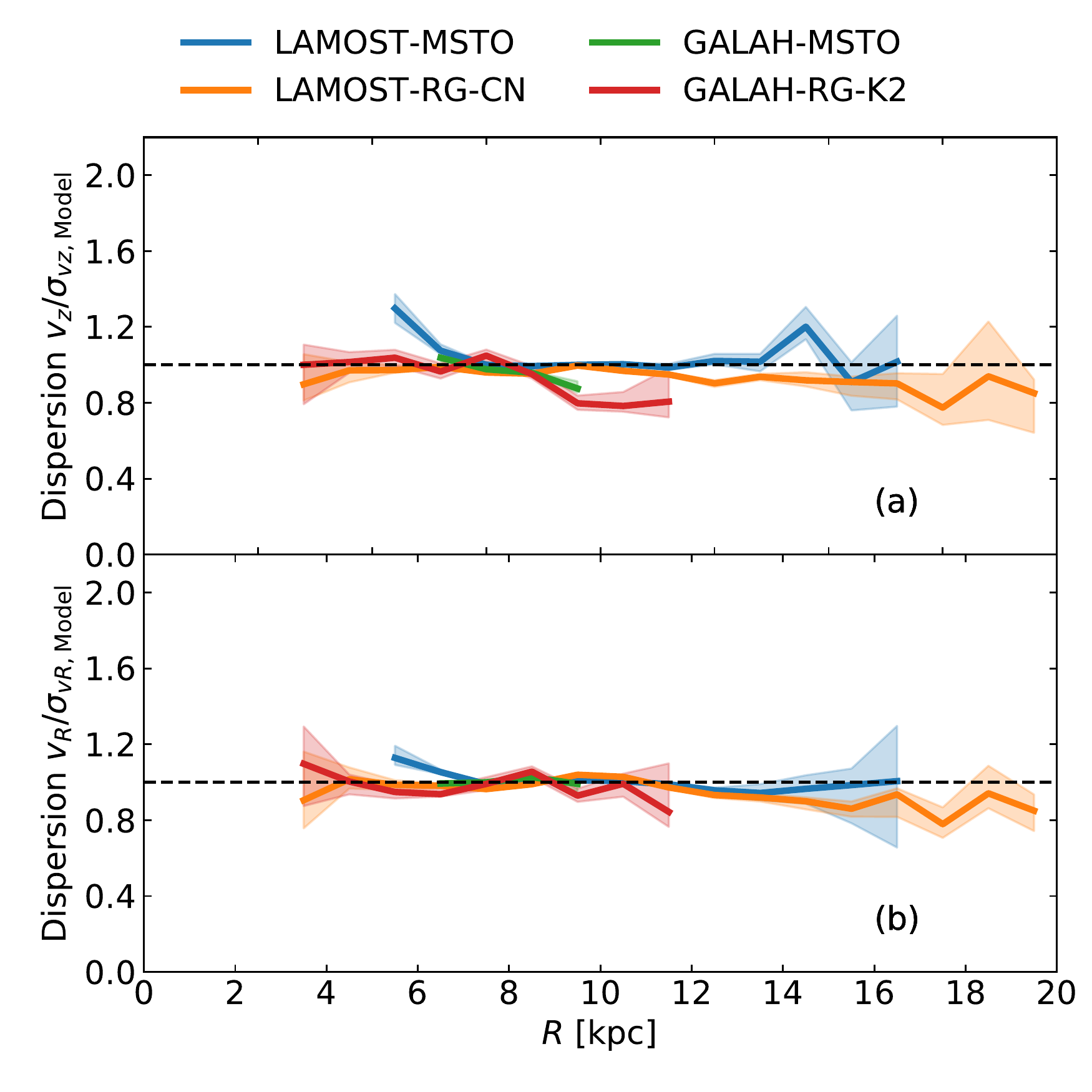}
\caption{Dispersion of normalized velocity as a function of Galactocentric radius $R$. The shaded region denotes the 16 and 84 percentile confidence interval estimated using bootstrapping.
The velocity is normalized by dividing with the velocity dispersion predicted by the model $\sigma_v(X,\theta_v)$ as given by \autoref{equ:vdisp_model}.
The dashed line corresponds to the expected value of 1 for the case where the model describes the data perfectly.
\label{fig:sigma_vz_vr_rgc}}
\end{figure}

\section{Method}
The dispersion $\sigma_v$ of velocity $v$ (for either $v_R$ or $v_z$), is assumed to depend on the stellar
age $\tau$, angular momentum $L_z$, metallicity [Fe/H], and vertical height from the disc midplane $z$, via the following multiplicatively separable functional form
\begin{equation}
\sigma_v(X,\theta_v)=\sigma_v(\tau,L_z,{\rm [Fe/H]},z,\theta_v)=\sigma_{0,v} f_{\tau}f_{L_z}f_{{\rm [Fe/H]}} f_{z}.
\label{equ:vdisp_model}
\end{equation}
Here, $X=\{\tau,L_z,{\rm [Fe/H]},z\}$ is a set of observables that are
independent variables and
\begin{equation}
f_{\tau}=\left(\frac{\tau/{\rm Gyr}+0.1}{10+0.1}\right)^{\beta_v},
\label{equ:f_tau}
\end{equation}
\begin{equation}
f_{L_z}=\frac{\alpha_{L,v} (L_z/L_{z,\odot})^2+\exp[-(L_z-L_{z,\odot})/\lambda_{L,v}]}{1+\alpha_{L,v}},
\label{equ:f_lz}
\end{equation}
\begin{equation}
f_{\rm [Fe/H]}=1+\gamma_{{\rm [Fe/H]},v} {\rm [Fe/H]},
\label{equ:f_feh}
\end{equation}
\begin{equation}
f_{z}=1+\gamma_{z,v} |z|,
\label{equ:f_z}
\end{equation}
and $\theta_v=\{\sigma_{0,v},\beta_{v},\lambda_{L,v},\alpha_{L,v},\gamma_{{\rm [Fe/H]},v},\gamma_{z,v}\}$ is a set of free parameters.
The functional forms were chosen based on a
preliminary analysis of the trends with respect to each observable.
The $\sigma_v$ has a power law dependence on age, with $\beta_{v}$ denoting the exponent. The age relation has a finite birth dispersion for stars younger than 0.1 Gyr.
The $\sigma_v$ falls off exponentially with $L_z$ with scale $\lambda_{L,v}$, but at large $L_z$ it is allowed to rise as $L_z^2$ (to account for flaring) and this rise is controlled by $\alpha_{L,v}$. The
$\sigma_v$ varies linearly with both [Fe/H] and $|z|$ with  gradients $\gamma_{{\rm [Fe/H]},v}$ and $\gamma_{z,v}$ respectively.
The $\sigma_{0,v}$ is a constant that
denotes the velocity dispersion for stars lying in the midplane with solar metallicity, solar angular momentum ($L_{z,\odot}=\Omega_{\odot}R_{\odot}^2$) and  an age of 10 Gyr.
The likelihood of the observed velocities $v=\{v_0,...v_N\}$ for a sample of $N$ stars can be written as
\begin{equation}
p(v_0,..v_N|X_0,...,X_N,\theta_v)  = \prod_i \mathcal{N}(v_i|0,\epsilon_{vi}^2+\sigma_v^2(X_i,\theta_v)),
\end{equation}
with $\epsilon_{vi}$ being the uncertainty corresponding to the observed velocity $v_i$ of the $i-$th star.
Here, $\mathcal{N}(v|\mu,\sigma^2)=\exp[-(v-\mu)^2/(2\sigma^2)]/\sqrt{2\pi\sigma^2}$ denotes the distribution of a random variable $v$ sampled from a  normal distribution
with mean $\mu$ and variance $\sigma^2$.
We find the maximum likelihood estimate (MLE) of $\theta_v$ by using the Nelder-Mead algorithm as implemented in the python package \texttt{scipy.optimize.minimize}. The MLE values of $\theta_v$ for the velocity components $v_z$ and $v_R$ are given in \autoref{tab:coeff}. Also given alongside are uncertainties, which were estimated
using bootstrapping.
The data used for estimating $\theta_v$
contained an equal number of stars
from the LAMOST and GALAH surveys.

To see the velocity dispersion profiles $f_{x}(x)$ that result from the observed data corresponding to each independent variable $x$, we must
factor out the dependence on other independent variables in the set $X$.
We accomplish this by binning
the stars in $x$ and computing the dispersion of $v/\sigma_{v,[x]}$
in each bin to get a profile of the dispersion as a function of
$x$.
Here, $\sigma_{v,[x]}$, defined as
\begin{equation}
\sigma_{v,[x]}(X,\theta_v)=\sigma_{0,v} \prod_{y \in X}^{x\neq y} f_y(y|\theta_v),
\label{equ:sigma_bracket}
\end{equation}
is the complementary velocity dispersion that includes all independent variables in set $X$ except $x$.

\begin{figure}
\centering \includegraphics[width=0.49\textwidth]{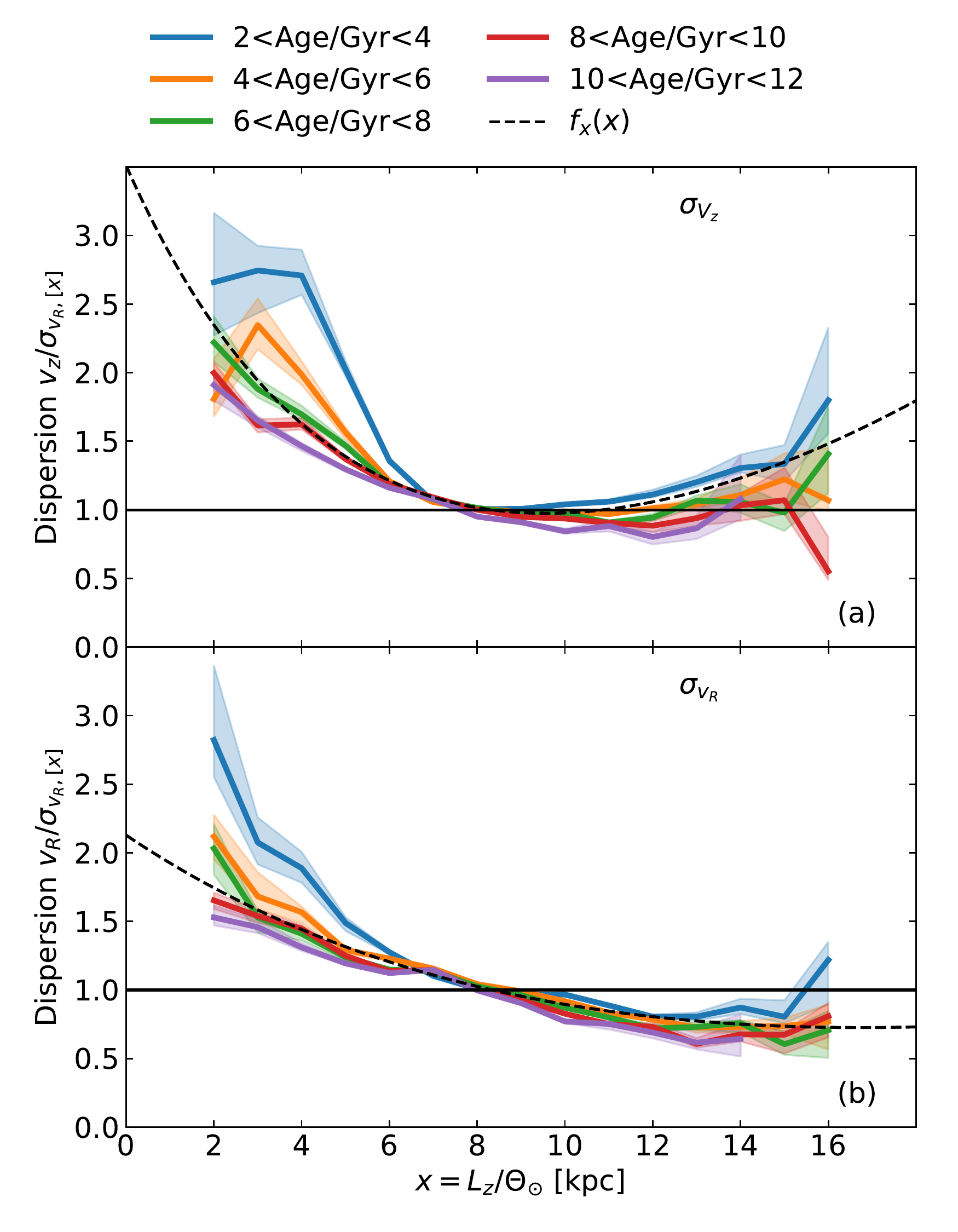}
\caption{Velocity dispersion as a function of angular momentum for stars lying in different age bins.
The shaded region denotes the 16 and 84 percentile confidence interval.
The dependence on other independent variables (age, [Fe/H] and $|z|$) have been factored out.  The observed stars are from LAMOST-MSTO, LAMOST-RG-CN, GALAH-MSTO, and GALAH-RG-K2 data sets.
The dashed lines show the the best fit model profiles.
The curves are normalized to have unit dispersion
at $L=L_{\odot}$.
The profiles do not show any significant variation with the age. The profile for the youngest bin is slightly steeper.
\label{fig:sigma_vz_vr_rg_age}}
\end{figure}

\begin{figure}
\centering \includegraphics[width=0.49\textwidth]{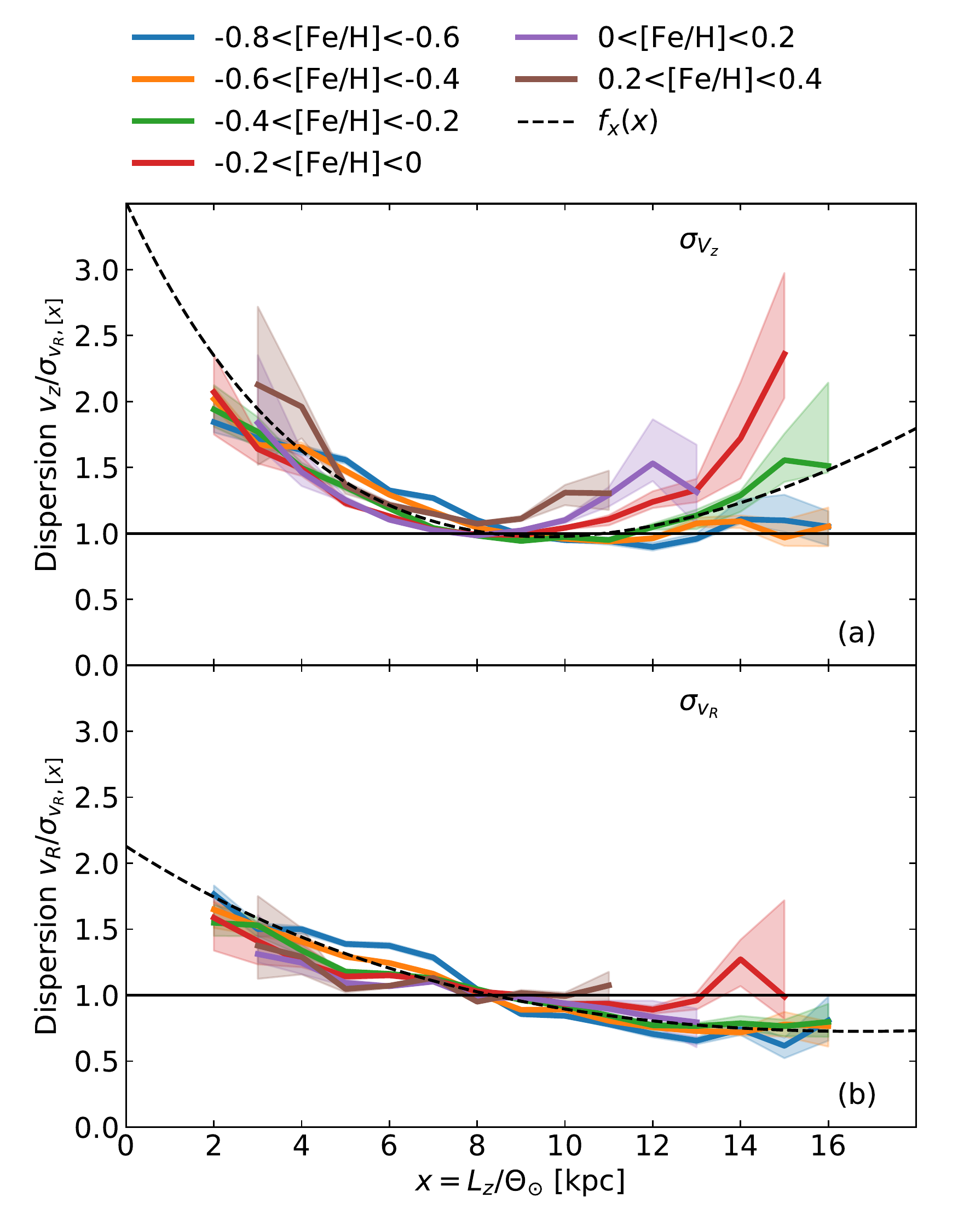}
\caption{Same as  \autoref{fig:sigma_vz_vr_rg_age} but for
stars lying in different [Fe/H] bins. The vertical dispersion is slightly higher for metal rich stars.
\label{fig:sigma_vz_vr_rg_fe_h}}
\end{figure}

\begin{figure}
\centering \includegraphics[width=0.49\textwidth]{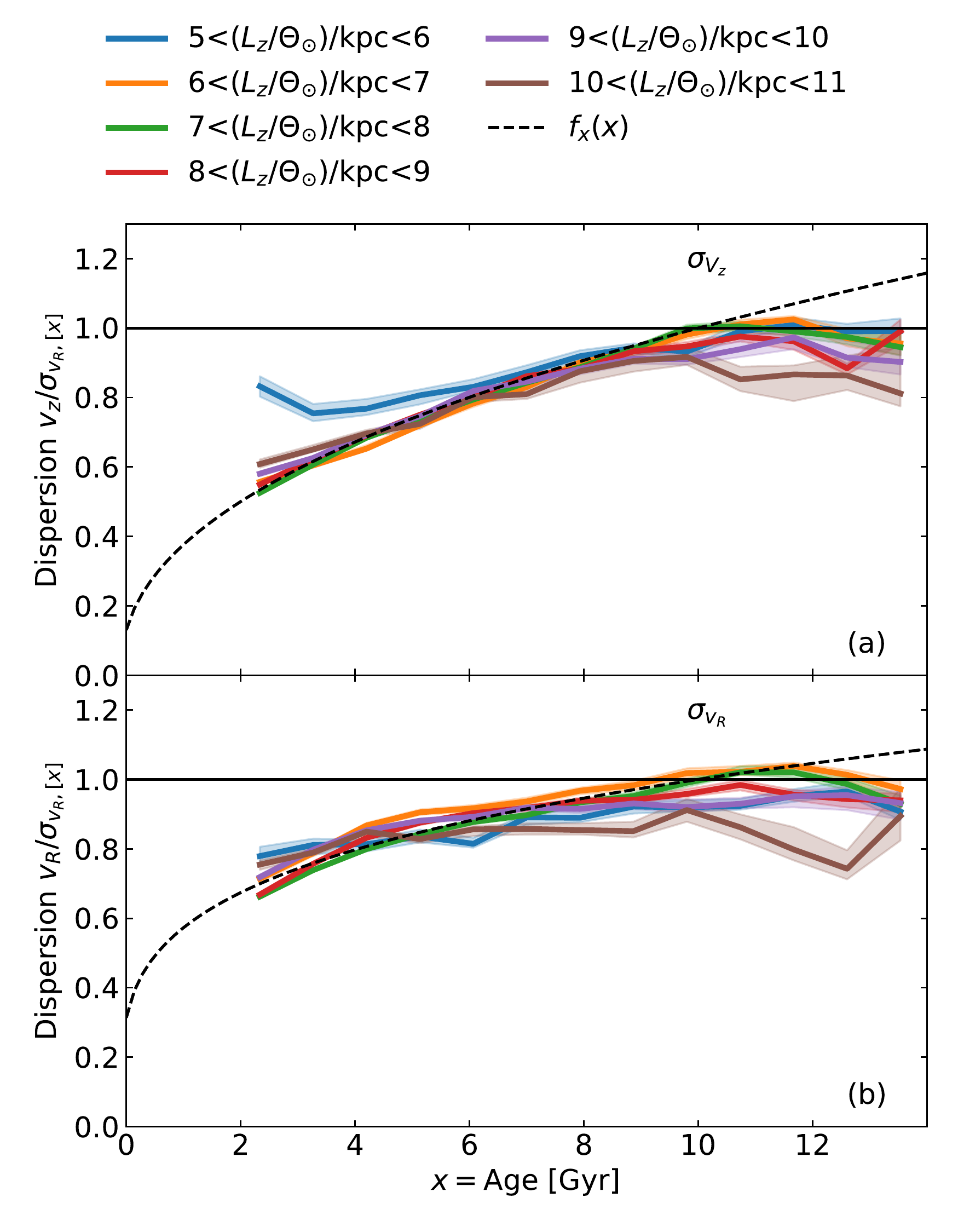}
\caption{Velocity dispersion as a function of age for stars lying in different $L_z$ bins. The shaded region denotes the 16 and 84 percentile confidence interval.
The curves are normalized to have unit dispersion
at an age of 10 Gyr.
The dependence on other independent variables ($L_z$, [Fe/H] and $|z|$) have been factored out. The description of the panels and the data set used are same as in \autoref{fig:sigma_vz_vr_rg_age}.
For older stars the profiles show a mild variation with $L_z$. \label{fig:sigma_vz_vr_age_rg}}
\end{figure}

\begin{figure}
\centering \includegraphics[width=0.49\textwidth]{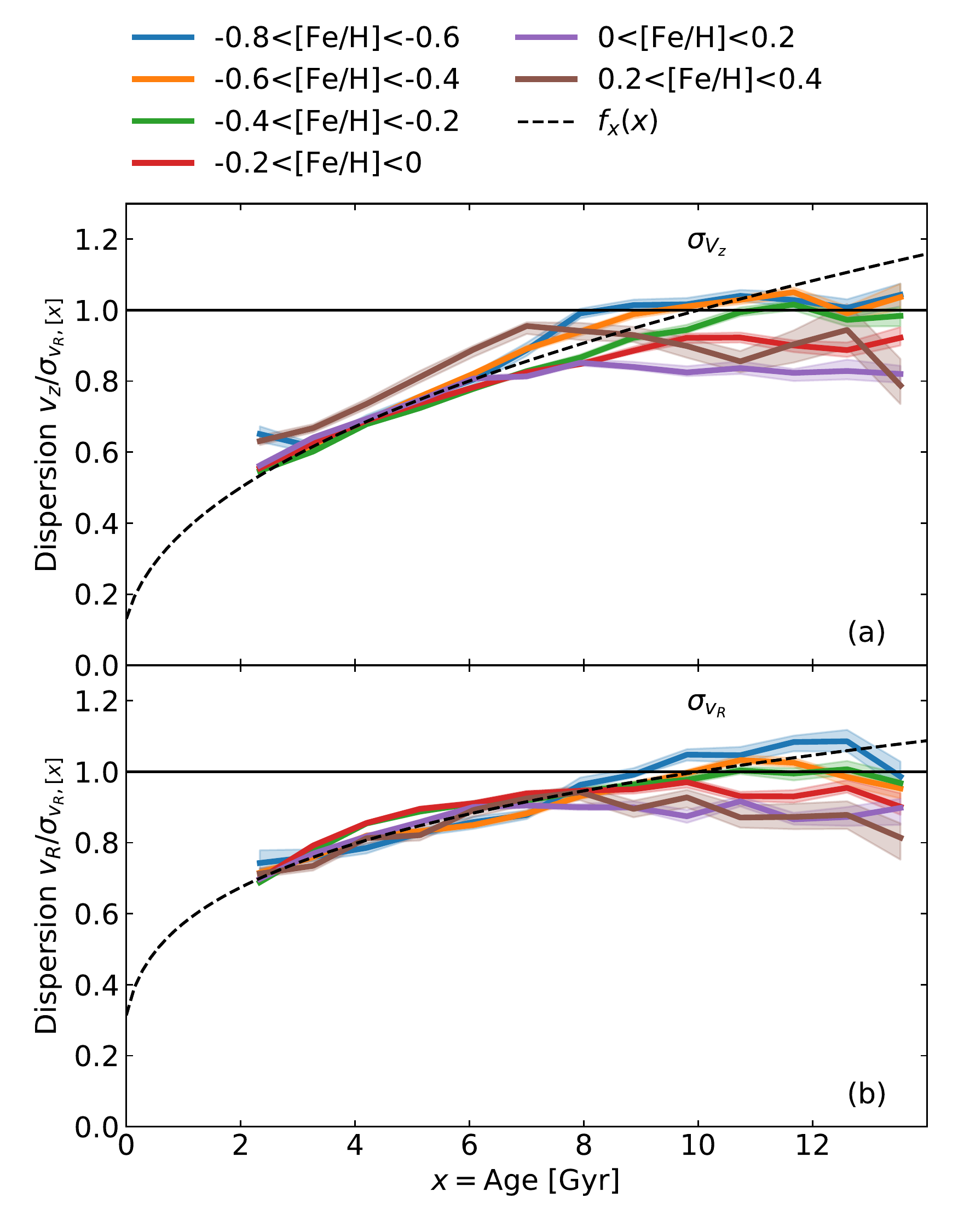}
\caption{Same as  \autoref{fig:sigma_vz_vr_age_rg} but for
stars lying in different [Fe/H] bins.
For older stars the profiles show a mild variation with [Fe/H].
\label{fig:sigma_vz_vr_age_fe_h}}
\end{figure}

\begin{figure}[tb]
\centering \includegraphics[width=0.49\textwidth]{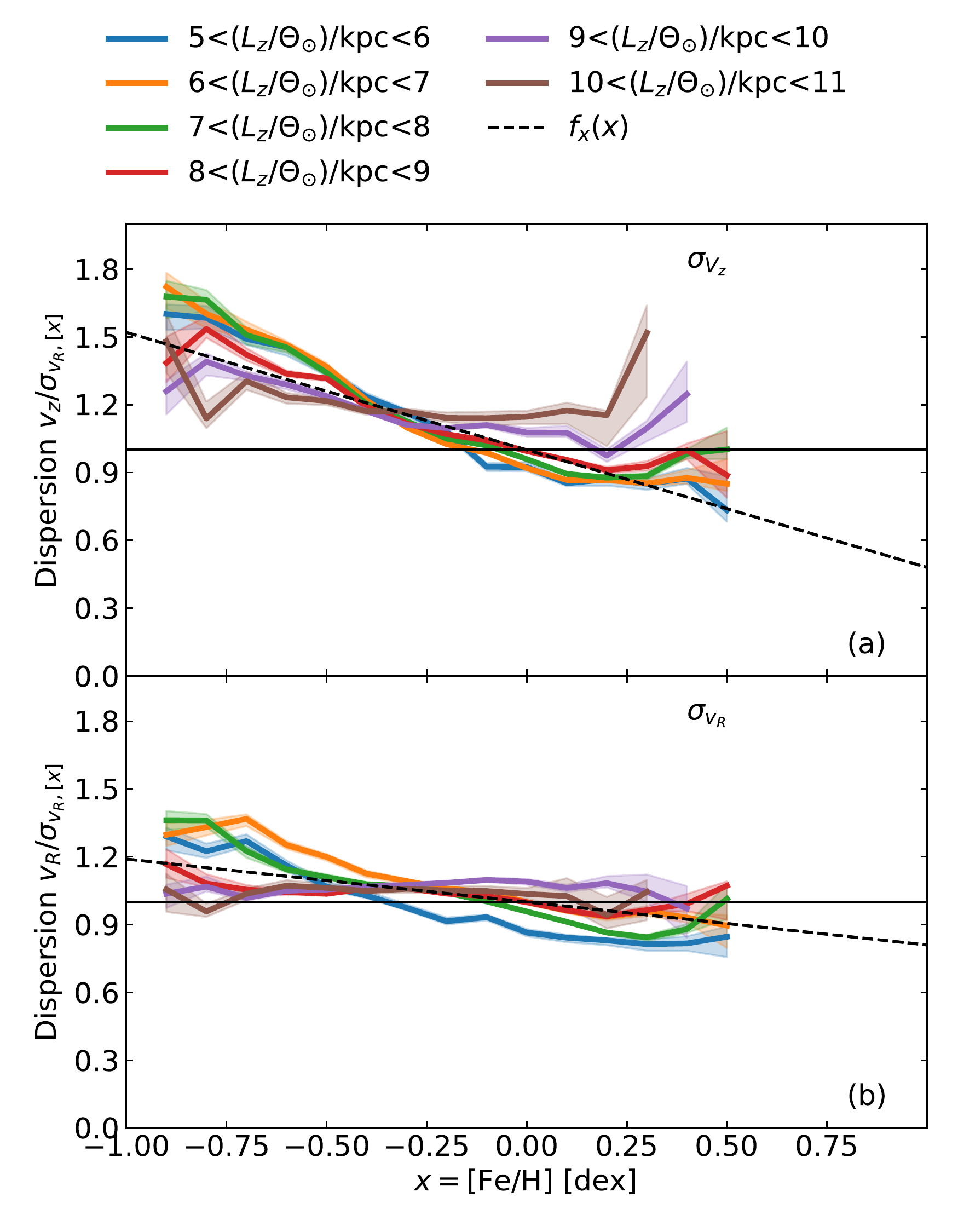}
\caption{
Velocity dispersion as a function of metallicity, [Fe/H], for stars lying in different angular momentum bins.
The shaded region denotes the 16 and 84 percentile confidence interval.
The curves are normalized to have unit dispersion
at [Fe/H]=0.
The dependence on other independent variables ($L_z$, age, and $|z|$) have been factored out. The description of the panels and the data set used are same as in \autoref{fig:sigma_vz_vr_rg_age}.
The profiles show a mild variation with $L_z$, with the profiles becoming flatter with increase of $L_z$.
\label{fig:sigma_vz_vr_fe_h_rg}}
\end{figure}

\begin{figure}[tb]
\centering \includegraphics[width=0.49\textwidth]{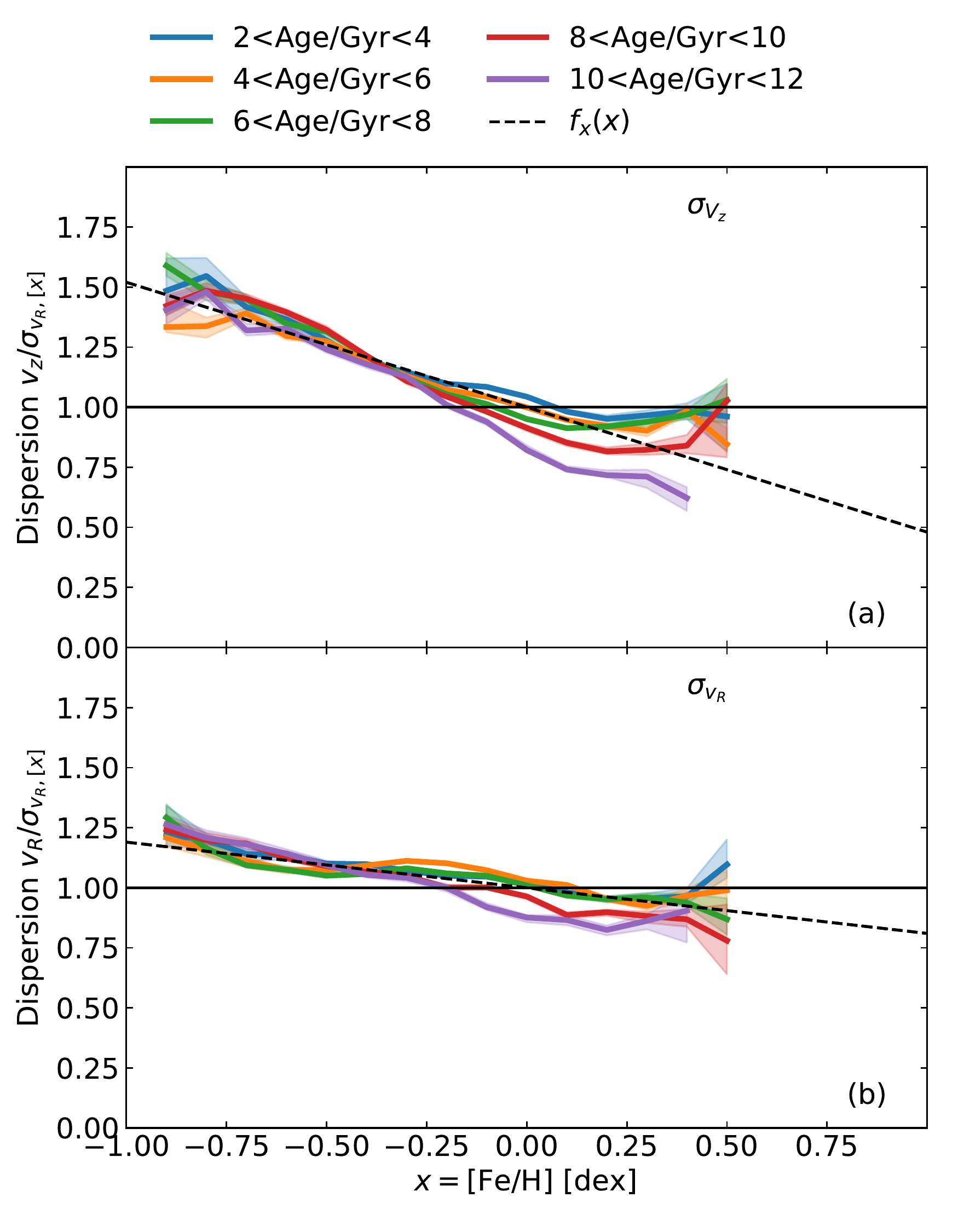}
\caption{Same as \autoref{fig:sigma_vz_vr_fe_h_rg} but for stars lying in different age bins. The profiles show a mild variation with the age, with the profiles becoming steeper with the increase of age.
\label{fig:sigma_vz_vr_fe_h_age}}
\end{figure}

\begin{figure}[tb]
\centering \includegraphics[width=0.49\textwidth]{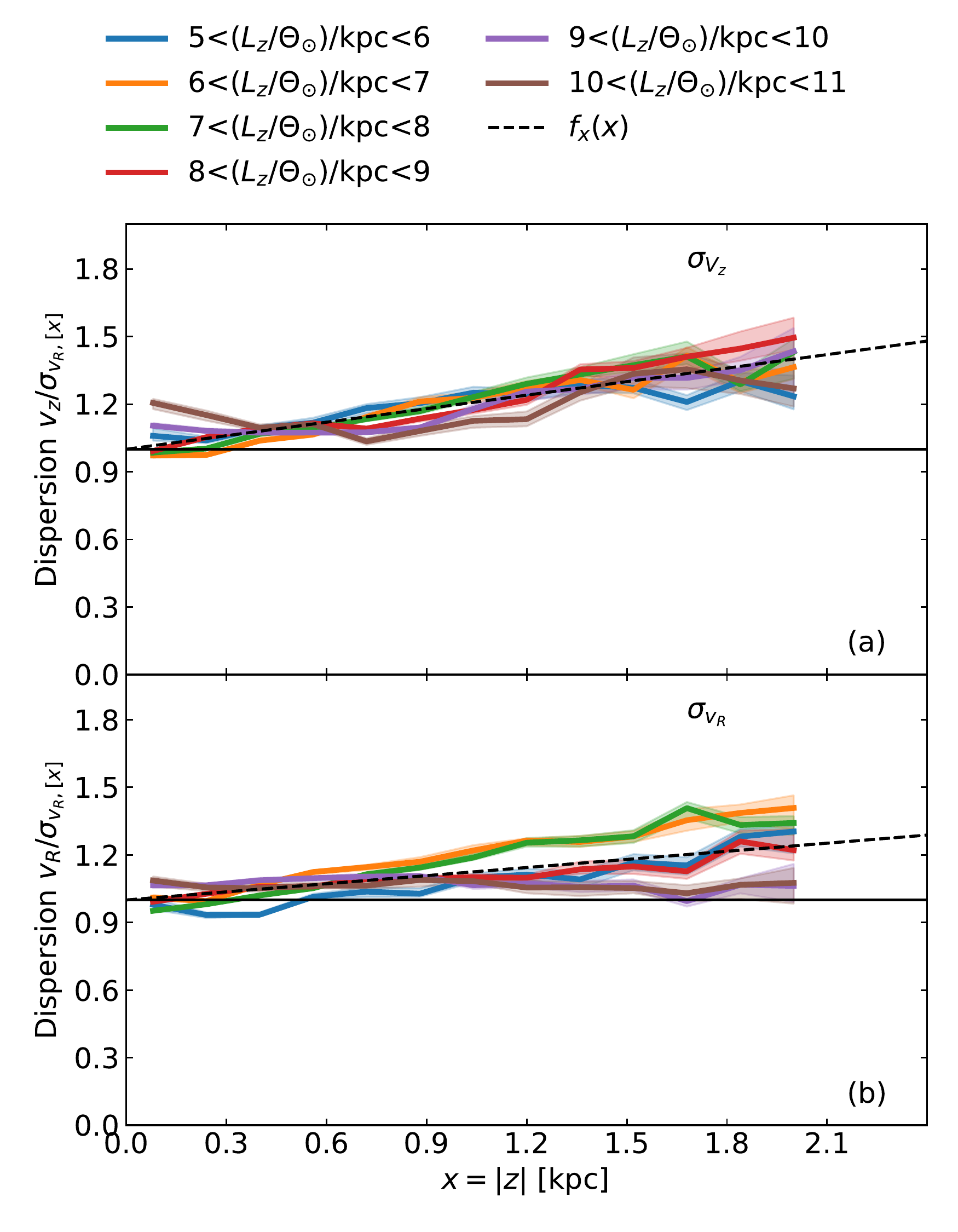}
\caption{Velocity dispersion as a function of distance $|z|$ from the mid-plane of the Galaxy, for stars lying in different angular momentum bins.
The shaded region denotes the 16 and 84 percentile confidence interval.
The dependence on other independent variables ($L_z$, age, and [Fe/H]) have been factored out.
The description of the panels and the data set used are same as in \autoref{fig:sigma_vz_vr_rg_age}.
The relationship for radial dispersion shows a mild variation with the $L_z$, with the slope becoming flatter with the increase of $Lz$.
\label{fig:sigma_vz_vr_pzgc_abs_rg}}
\end{figure}

\begin{figure}[tb]
\centering \includegraphics[width=0.49\textwidth]{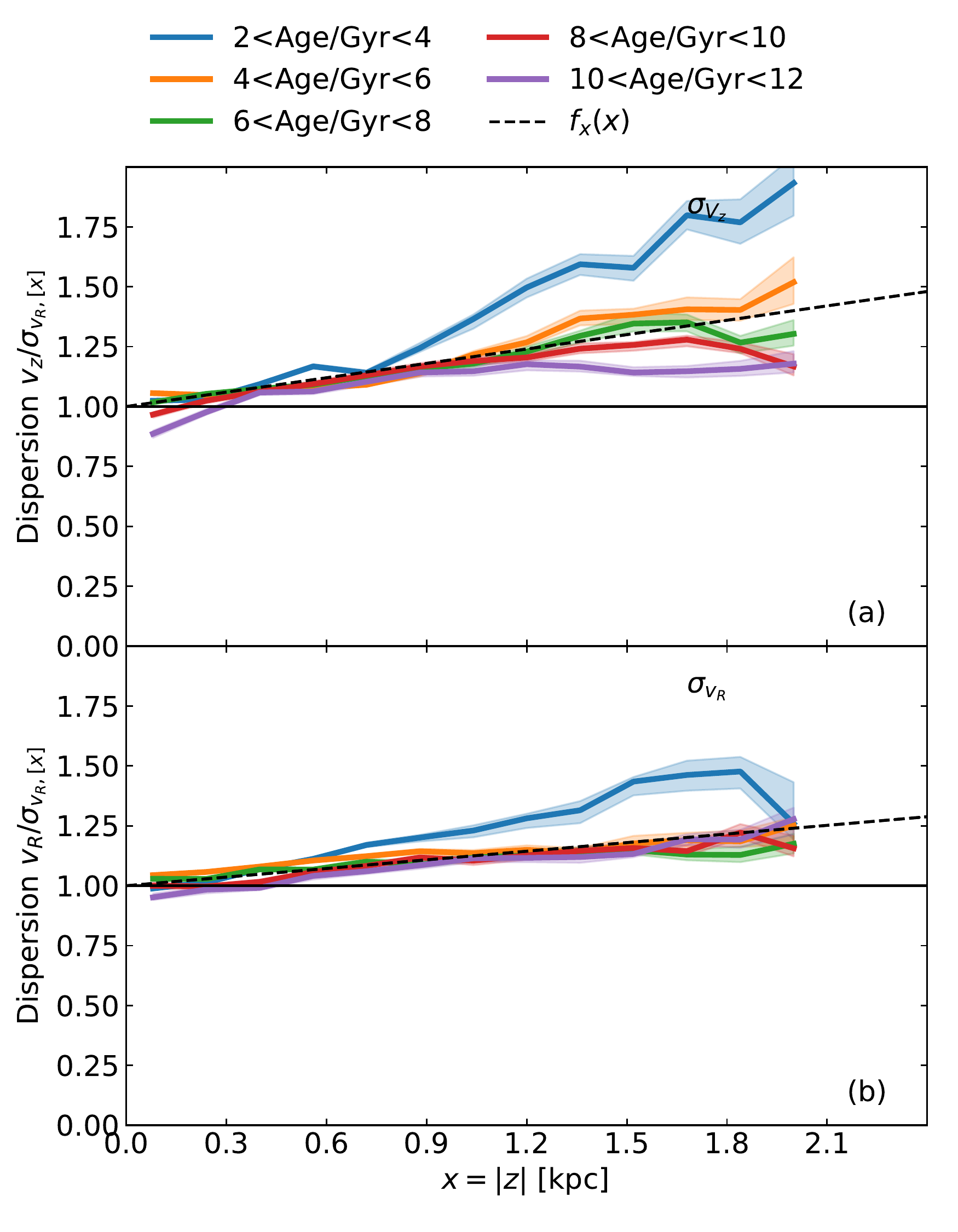}
\caption{Same as \autoref{fig:sigma_vz_vr_pzgc_abs_rg}
but for stars lying in different age bins. The slope is higher for younger stars.
\label{fig:sigma_vz_vr_pzgc_abs_age}}
\end{figure}

\begin{figure}
\centering \includegraphics[width=0.49\textwidth]{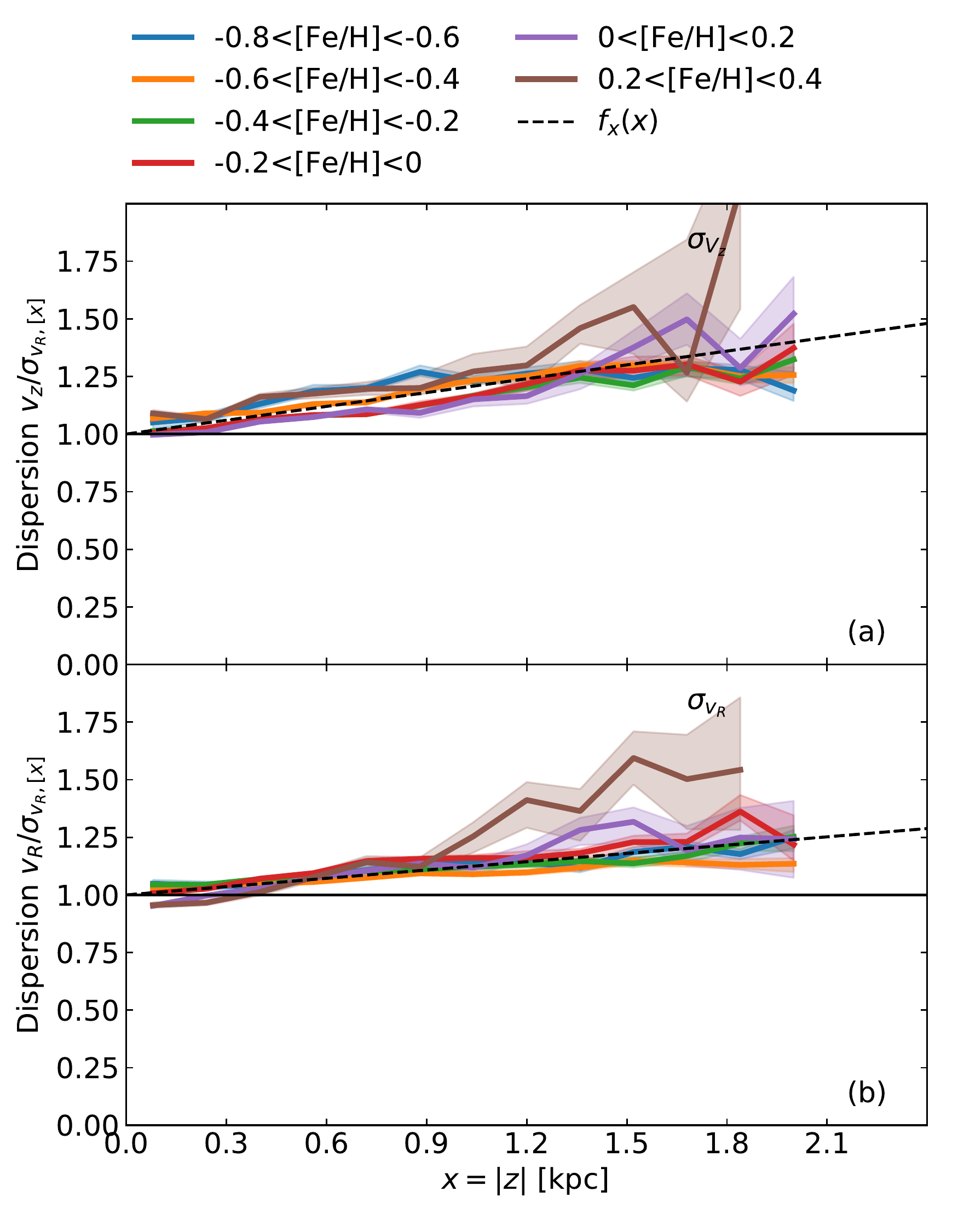}
\caption{Same as \autoref{fig:sigma_vz_vr_pzgc_abs_rg}
but for stars lying in different [Fe/H] bins. The relationship shows
very little variation with [Fe/H].
\label{fig:sigma_vz_vr_pzgc_abs_fe_h}}
\end{figure}

\section{Results}
\subsection{Basic trends}
Our basic trends with angular momentum, age, metallicity and perpendicular distance from the plane are shown in \autoref{fig:sigma_vz_vr}. The top panels explore
the vertical dispersion while the bottom panels explore the radial dispersion. The dispersion is assumed to be a function of multiple independent variables, and from left to right each panel shows
the effect of one independent variable at a time. The dispersions
decrease exponentially with $L_z$ up to about the solar angular momentum,
beyond that $\sigma_z$ starts to increase (\autoref{fig:sigma_vz_vr}a) whereas $\sigma_R$ flattens (\autoref{fig:sigma_vz_vr}e).
The dispersions increase with age and are well described by a power law, with the exponent
being higher for $\sigma_z$ as compared to $\sigma_R$ (\autoref{fig:sigma_vz_vr}b and \autoref{fig:sigma_vz_vr}f). The GALAH-MSTO
data set does not show any saturation for older stars, but other data sets
show a flattening for stars older than 10 Gyr. The dispersions decrease
linearly with both [Fe/H] (\autoref{fig:sigma_vz_vr}c and \autoref{fig:sigma_vz_vr}g) and $|z|$ (\autoref{fig:sigma_vz_vr}d and \autoref{fig:sigma_vz_vr}h), with the slope being steeper for
$\sigma_z$. \autoref{fig:sigma_vz_vr_rgc} demonstrates that
all of the above discussed trends are independent of the location $R$
and are valid for $3<R/{\rm kpc}<20$.

Data from different surveys and stellar types are shown separately in
\autoref{fig:sigma_vz_vr}. \autoref{fig:sigma_vz_vr}a and \autoref{fig:sigma_vz_vr}b show some systematic differences, but overall the different data sets are all found to be consistent with the same relationship.
Agreement between the GALAH and LAMOST results suggests that their spectroscopic parameters do not not have any strong systematics with respect to each other. Agreement between data sets of different stellar types (MSTO and RG), suggests there are no strong systematics related to stellar types. This is very reassuring and useful given the fact that for different stellar types very different age estimation techniques are used.

\subsection{Variation of basic trends with respect to other independent variables}
In \autoref{fig:sigma_vz_vr_rg_age} to
\autoref{fig:sigma_vz_vr_pzgc_abs_fe_h},
we show the residual dependence of each relation on other independent variables. In each figure, we plot the
observed relation corresponding to one independent variable, such as the relation $f_{\tau}$, by binning stars in another independent variable, e.g., $L_z$ or [Fe/H].
In general we find that the relations vary
very little with respect to other independent variables, suggesting
that modelling the dispersion as a product of multiple independent functions, as given by \autoref{equ:vdisp_model},
is a good approximation. However, slight variations can be seen. Interestingly, the variations that
we see are in most cases systematic and we discuss these
systematic trends below.

The $f_{L_z}$ relation varies with age such that the younger stars have higher relative dispersion (\autoref{fig:sigma_vz_vr_rg_age}).
Here by relative dispersion we mean dispersion relative to the derived relations.
The $f_{L_z}$ relation varies with [Fe/H], such that for high $L_z$, the metal rich stars have systematically higher relative dispersion (\autoref{fig:sigma_vz_vr_rg_fe_h}).
The $f_{\tau}$ relation does not vary for stars
with age less than 8 Gyr, but for older stars the relative dispersion is systematically lower for the high-$L_z$ stars (\autoref{fig:sigma_vz_vr_age_rg}) and for high-metallicity stars (\autoref{fig:sigma_vz_vr_age_fe_h}). This could be because both  old high-$L_z$ stars and old high-metallicity stars are rare and hence
an old star bin is more likely to be contaminated
by young stars (due to age uncertainties), which will lower the overall dispersion in that bin given that young stars have low dispersion.

The $f_{\rm [Fe/H]}$ relation flattens with increasing $L_z$ (\autoref{fig:sigma_vz_vr_fe_h_rg}). The relation
does not vary much with age, with the exception that for old stars
the relative dispersion is lower at the high metallicity end.
(\autoref{fig:sigma_vz_vr_fe_h_age}).
This again could be due to old and high-metallicity stars being rare
and hence an old star bin is more likely to be contaminated
by young stars.
The $f_{z}$ relation for vertical velocity dispersion does not seem to vary much with $L_z$ (\autoref{fig:sigma_vz_vr_pzgc_abs_rg}a), however, the $f_{z}$ relation for radial velocity dispersion flattens with increase in $L_z$ (\autoref{fig:sigma_vz_vr_pzgc_abs_rg}b).
The $f_{z}$ relation does not vary much with age (\autoref{fig:sigma_vz_vr_pzgc_abs_age}) or
[Fe/H] (\autoref{fig:sigma_vz_vr_pzgc_abs_fe_h}). However, for stars younger than 4 Gyr the relationship is much steeper (\autoref{fig:sigma_vz_vr_pzgc_abs_age}).

\subsection{Systematics between surveys}
\begin{figure}[tb]
\centering \includegraphics[width=0.49\textwidth]{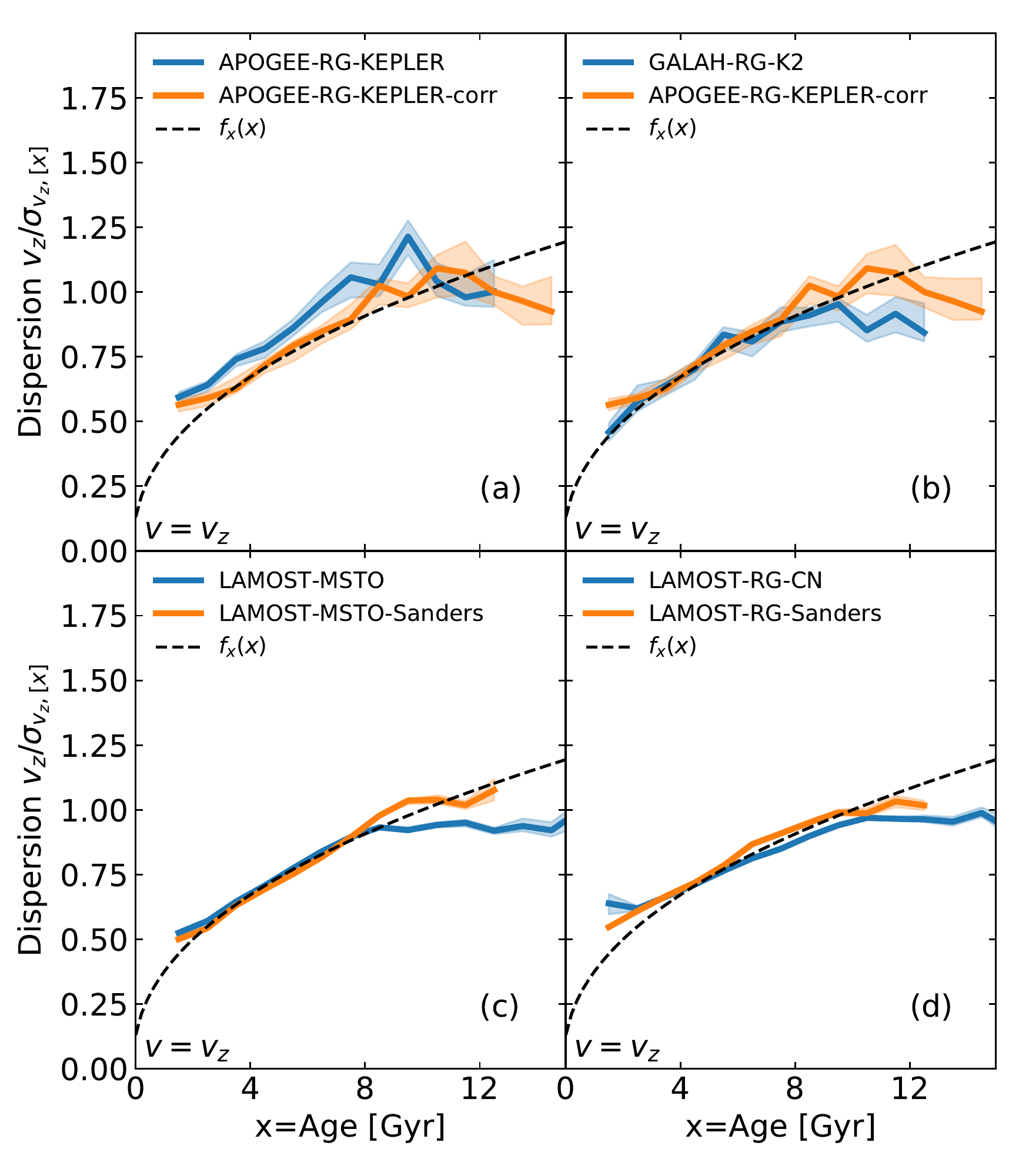}
\caption{Velocity dispersion as a function of age for different data sets. The 16 and 84 percentile confidence interval is denoted by the shaded region.
The dependence on other independent variables ($L_z$, [Fe/H], and $|z|$) have been factored out.
The data set APOGEE-RG-KEPLER-corr was generated from
APOGEE-RG-KEPLER by correcting the metallicity and age by the following transformations ${\rm [Fe/H]_{corr}=[Fe/H]-0.1}$ and  $\tau_{\rm corr}=1.1 \tau$. The velocity dispersion seems to saturate
when using ages from the LAMOST value added catalog. In comparison, no such saturation is seen when using ages from the \citet{2018MNRAS.481.4093S} catalog.
\label{fig:sigma_vz_vr_age_kepler_sanders}}
\end{figure}

In \autoref{fig:sigma_vz_vr} we presented results from four different data sets.
Two were based on the LAMOST spectroscopic survey, with one made up of MSTO stars and other made up of RGB stars.
The other two data sets were based on the GALAH spectroscopic survey, with one made up of MSTO stars and other of RG  stars having asteroseismic information from K2.
Another large spectroscopic survey that we did not use in \autoref{fig:sigma_vz_vr} was APOGEE.
In \autoref{fig:sigma_vz_vr_age_kepler_sanders}a, we plot
APOGEE results for the asteroseismic sample from {\it Kepler}
\citep{2018ApJS..239...32P}.
It shows that the observed AVR is shifted with respect
to our empirical relations.
Examination of stars in between the
APOGEE and GALAH/LAMOST data sets, revealed that the APOGEE iron abundances were
systematically higher by 0.1\,dex. However, this is still
not enough to account for the difference seen in \autoref{fig:sigma_vz_vr_age_kepler_sanders}a. A further increase in the age of APOGEE stars by 10\%
is required to bring the the APOGEE-RG-KEPLER data set into agreement with the other datasets.
These systematic offsets are the reason  APOGEE data was not used in the analysis presented in \autoref{fig:sigma_vz_vr}.
\autoref{fig:sigma_vz_vr_age_kepler_sanders}b shows that with these
changes the APOGEE-RG-KEPLER data set can also be brought into agreement with
the GALAH-RG-K2 data.

In \autoref{fig:sigma_vz_vr}, the LAMOST-MSTO and LAMOST-RG-CN data sets show significant flattening of the AVR for age greater than 10 Gyr,
while such a flattening is not seen for the GALAH-MSTO stars.
The flattening for LAMOST data sets could be due to larger
uncertainties on age estimates in them.
\autoref{fig:sigma_vz_vr_age_kepler_sanders}c shows results
for the LAMOST-MSTO and LAMOST-RG-CN data sets, using
age estimates from \citet{2018MNRAS.481.4093S}. No flattening is
seen here.
This could be because the \citet{2018MNRAS.481.4093S} ages are more precise than LAMOST ages for the older stars. However, it should be noted that \citet{2018MNRAS.481.4093S} use strong priors based on height above the plane,
which can increase the age precision for older stars, but is not ideal to study trends with height above the plane, as we do in this paper.

\subsection{The role of angular momentum in shaping the solar-neighborhood AVR}
\begin{figure}[tb]
\centering \includegraphics[width=0.49\textwidth]{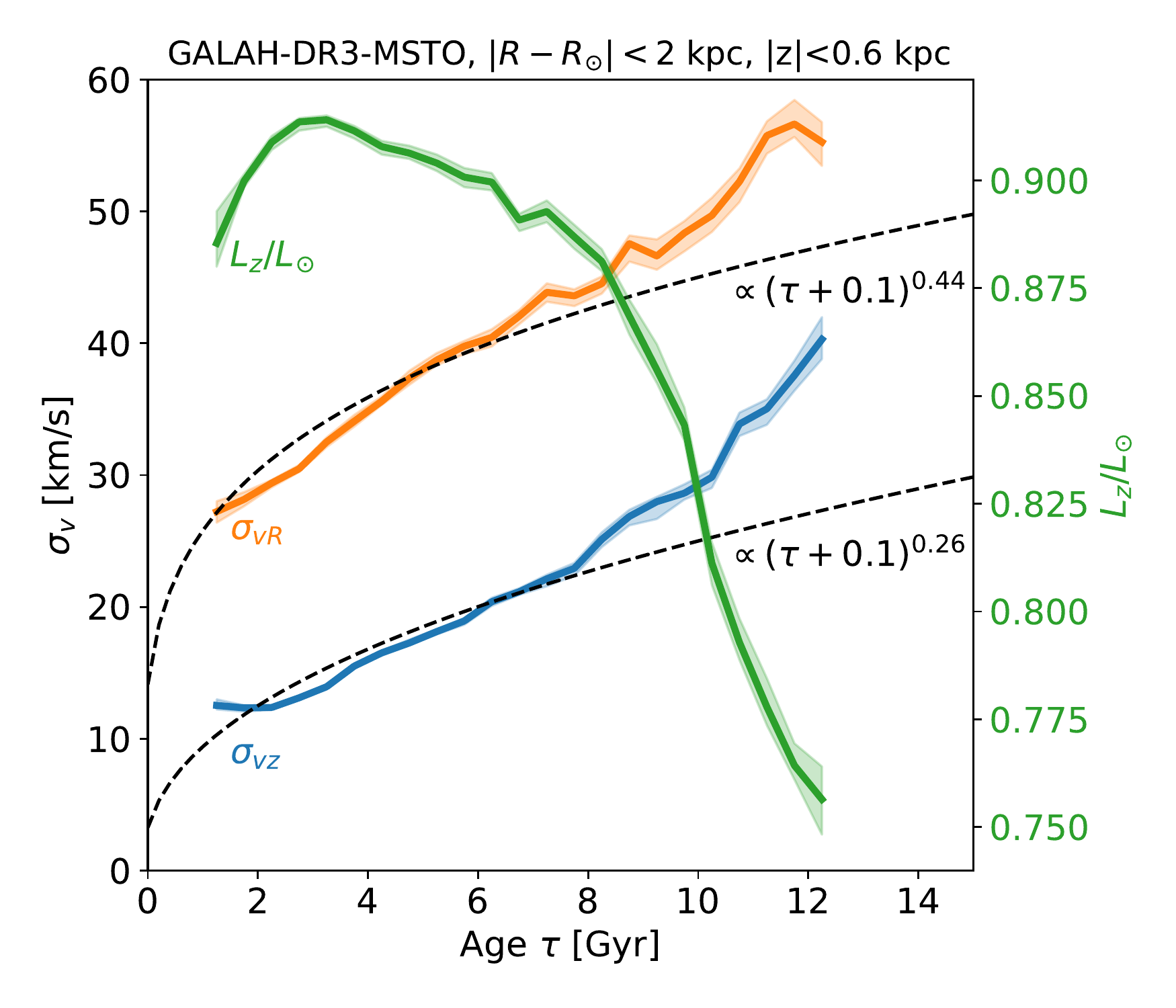}
\caption{Velocity dispersion as a function of age for stars in the solar neighborhood.
Plotted alongside
is median angular momentum as a function of age.
Dashed lines show power law profiles. For age greater than 8 Gyr, the velocity dispersion breaks away from the plotted power law profiles, and this break
coincides with a fall in angular momentum.
\label{fig:sigma_vz_vr_age_galah}}
\end{figure}

Angular momentum plays a critical role in shaping the
AVR of stars observed in the solar neighborhood. It has been claimed
in some previous studies that the AVR deviates from a power law with an abrupt increase for old stars \citep{1991dodg.conf...15F, 1993A&A...275..101E, 2001ASPC..230...87Q}.
Since, most old stars in the solar neighborhood belong to the thick disc, this tentatively suggests that the kinematics of the thick disc stars is different from that of the thin disc stars.
We show in \autoref{fig:sigma_vz_vr_age_galah} that this
apparent break is due to a systematic variation of angular momentum with age, since
older stars have low angular momentum and low angular momentum stars
have higher velocity dispersion.
Once again, when a variation of angular momentum is allowed for, all stars seem to be consistent with a universal AVR.
However, it is still not clear as to why the angular momentum decreases with age. It could be due to inside out formation
of the disc and radial migration of stars from the
inner disc and needs to be investigated in future.

\section{Discussions}
\subsection{ Dependence of dispersions on age}
We find $\beta_z=0.44$ and $\beta_R=0.26$, which is in good agreement with predictions of simulations by \citet{2016MNRAS.462.1697A}, where the effects of spiral arms, GMCs and a bar is taken into account.
For the GALAH-MSTO stars, even the old thick disc stars satisfy this relationship. The LAMOST-MSTO, LAMOST-RG-CN and GALAH-RG-K2 all show saturation for age greater than 10 Gyr. This could be due to significant uncertainties in ages for the older stars
in samples other than GALAH-MSTO. For example, the uncertainty of asteroseismic ages is known to
increase with age and is predicted to be around 30\% for RGB stars and even higher for red-clump stars \citep{2018MNRAS.475.5487S}. Since old stars are
typically rare, an old star bin is more likely to be contaminated
by young stars (due to age uncertainties), which will lower the overall dispersion in that bin given that young stars have low dispersion.

As discussed in \citet{2016MNRAS.462.1697A}, the exponents $\beta_z$ and $\beta_R$ of the AVR
depend  in a complex way upon the whole dynamical history of the Galaxy. This is because there are at least two major scattering agents (spiral structures and GMCs) and the strength of scattering due to them changes with time.
Spiral structure is mainly responsible for in-plane scattering, while GMCs contribute to both in-plane and vertical scattering. Spiral structure drives up $\sigma_R$ fairly rapidly,   increasing the Toomre stability paramater $Q$ and making the disc stable. This makes the ratio $\sigma_z/\sigma_R$ very small initially.
Thereafter, on a longer time scale, scattering from GMCs increases both $\sigma_R$ and $\sigma_z$. For scattering
from stationary fluctuations the exponent $\beta$ is predicted to be around 0.25, with $\beta_z$ being slightly higher than $\beta_R$ \citep{2002MNRAS.337..731H}. This is seen in the heating history of coeval populations. However, $\beta_z$ for the AVR is much higher
because the overall efficiency of heating due to GMCs
is higher at earlier times \citep[Figure-7 of][]{2016MNRAS.462.1697A}.  At earlier times, the star formation rate is high and the stellar disc mass is low and this makes the GMC mass fraction higher, which in turn increases the efficiency of GMC scattering.

\subsection{Dependence of dispersions on angular momentum} \label{sec:dep_lz}
Without any loss of generality, in what follows, we discuss our results in terms of guiding radius rather than angular momentum, with the guiding radius being defined as $R_g=L_z/\Theta_{\odot}$ (angular momentum divided by circular velocity at Sun).
We find that for $R_g<R_{\odot}$ both dispersions fall off exponentially with $R_g$.
In general, the strength of the secular heating processes, like
those due to spiral arms or GMCs, are expected to be proportional
to the surface density of stars $\Sigma$, so the dispersion
is expected to fall off with radius. Using some simple
physically motivated arguments we now predict the
radial scale length $R_{\sigma}$ of the exponential fall of dispersion
with $R$.
The vertical dispersion is expected to vary with surface density $\Sigma$, and scale height $h_z$, as $\sigma_z \propto \sqrt{\Sigma h_z}$ \citep{1988A&A...192..117V}. If $h_z$ is constant then the scale length of vertical dispersion, $R_{\sigma_z}$, should be
related to the scale length of stellar surface density, $R_d$, as $R_{\sigma_z}=2 R_d$. Using this we estimate $R_{\sigma_z}=\lambda_{L,vz}/\Theta_{\odot}=4.9 \pm 0.2$ kpc \citep[adopting $\Theta_{\odot}=232$ km/s from][]{2014ApJ...793...51S},
which is in good  agreement with the theoretical prediction of 5.0 kpc
\citep[adopting $R_d=2.5$ kpc, see e.g.,][]{2003A&A...409..523R,2008ApJ...673..864J}.
For the radial dispersion we expect, $\sigma_R \propto \Sigma R$ and hence $R_{\sigma_R}=R_d$. This follows from assuming the Toomre stability parameter
\be
Q=\frac{\sigma_R\kappa}{3.36{\rm G}\Sigma}
\ee
to be constant throughout the disc and the rotation curve to be flat, which implies $\kappa \propto 1/R$.  However we find $R_{\sigma_R}$ to be 9.9 kpc, which is about four times larger than $R_d$ (assuming $R_d=2.5$ kpc). Hence, the
observed scale length of $\sigma_R$ cannot be explained by a constant $Q$.

For $R_g >  R_{\odot}$ kpc, both dispersions break away from being purely exponential functions of $R_g$, with $\sigma_z$ increasing
and $\sigma_R$ flattening with $R_g$. This is indicative of flaring in the outer disc for mono-age and mono-metallicity populations.
Now, outside the solar radius
significant flaring has been reported for all mono age populations, with the flaring being strongest for the youngest population \citep{2017MNRAS.471.3057M}. Flaring implies  that $\sigma_z$ should fall off slower than $\sqrt{\Sigma}$, be constant, or rise. Hence, our overall reported rise of $\sigma_z$ with $R_g$ in the outer disc, and also the fact that the rise is stronger for younger stars (\autoref{fig:sigma_vz_vr_rg_age}) is consistent with the findings of \citet{2017MNRAS.471.3057M} related to flaring.

The flattening of velocity dispersion with $R_g$  is easy to understand.
Stars are thought to be born out of the inter-stellar medium with a
birth dispersion of about 10\,km/s,
due to turbulence in the medium driven by the injection of energy from newly forming stars.
Due to this non-zero lower bound on
the dispersion of newly forming stars, at large $R_g$
the dispersion cannot keep on falling exponentially but
will hit a floor and flatten.

We see in \autoref{fig:sigma_vz_vr}b and \autoref{fig:sigma_vz_vr}d  that the flattening occurs at a smaller value of $R_g$ for
$\sigma_z$ as compared to $\sigma_R$. This is  also
easy to understand.
Both $\sigma_z$  and
$\sigma_R$ are exponential functions
of $R_g$, but the overall proportionality constant for $\sigma_R$
is larger than $\sigma_z$. Additionally, the scale length for
$\sigma_R$ is larger than that for $\sigma_z$.
Consequently, flattening due to a constant birth dispersion will set in at a lower
value of $R_g$ for $\sigma_z$ than for $\sigma_R$.
The youngest stars also have the lower overall
proportionality constant for dispersion, so they are expected to
flatten earlier and this is visible in \autoref{fig:sigma_vz_vr_rg_age}.
The fact that non-zero birth
dispersion can lead to flattening
of dispersions and consequently flaring
has been nicely demonstrated by \citet{2016MNRAS.459.3326A}
using N-body simulations of discs
having spiral arms, GMCs and a bar.

What causes the dispersions to rise for $\sigma_z$
and why doesn't it also rise for $\sigma_R$?
This needs to be investigated in future. Simulations by \citet{2016MNRAS.462.1697A},
incorporating the effects of spiral perturbations, a bar, and GMCs,
only predict a monotonic fall or flattening for $\sigma_z$, but no rise
of dispersion with radius (see their Figure 4).
This suggests that some additional processes might be at play.
For example, the interaction of the disc with orbiting satellites \citep{2008ApJ...688..254K,2008MNRAS.391.1806V,2009ApJ...707L...1B} is known to cause flaring in the outer disc.
The infall of misalinged gas
\citep{2010MNRAS.408..783R,2012ApJ...750..107S,2013MNRAS.434.3142A} and reorientation of the  disc axis \citep{2013MNRAS.428.1055A} is also known to cause warps and consequently flaring.
Interestingly, one of the disc galaxies (Au18) simulated in a  cosmological context
by \citet{2016MNRAS.459..199G} shows a rise of $\sigma_z$ with $R$.
Since, this simulation does not have GMCs,
\citet{2016MNRAS.459..199G} attribute
the vertical heating to the bar and the effect of orbiting
satellites.

\begin{figure}[tb]
\centering \includegraphics[width=0.49\textwidth]{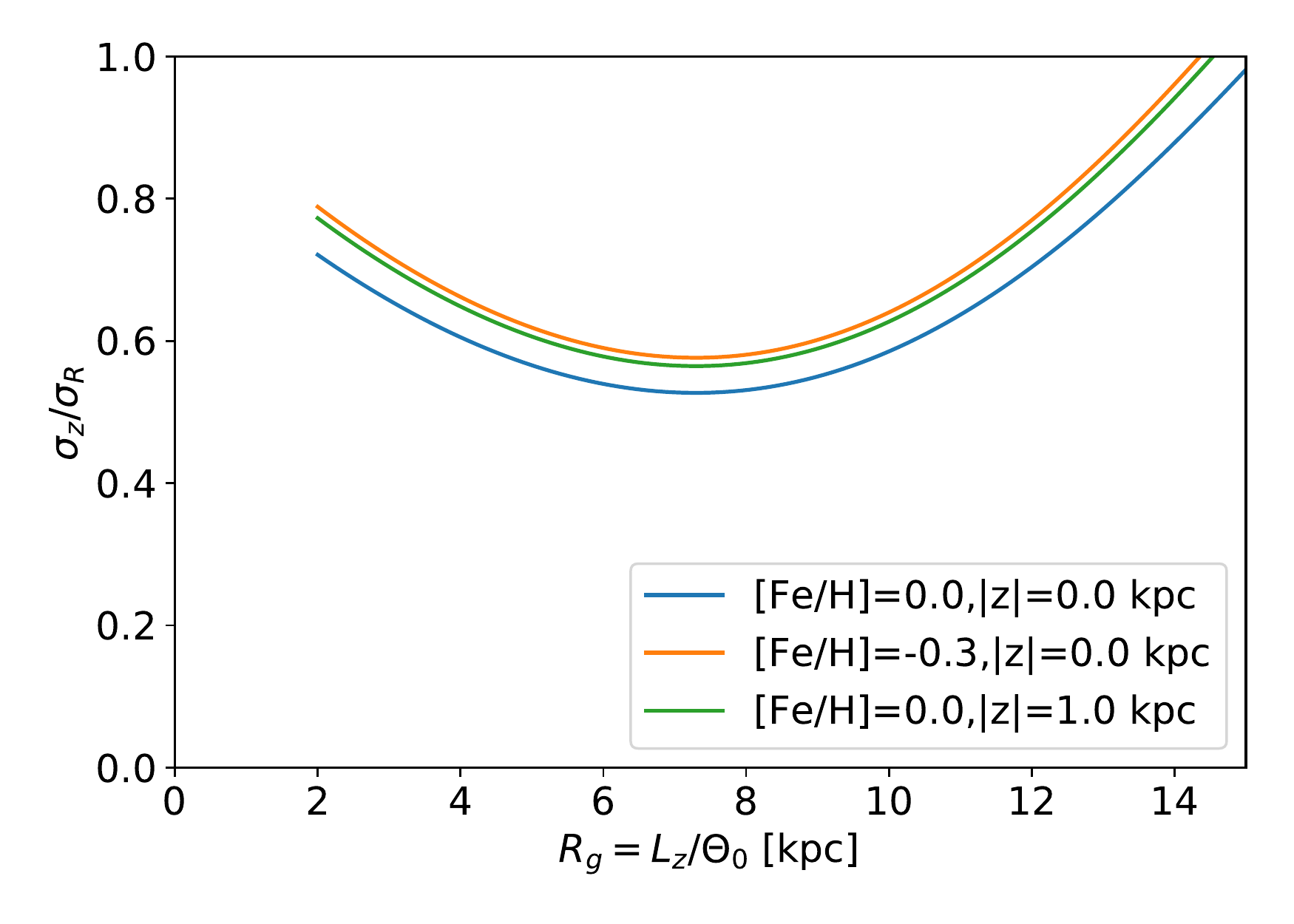}
\caption{The ratio $\sigma_z/\sigma_R$ as a function of guiding radius for stars of different age and metallicity.
The ratio is estimated using the analytical model described by \autoref{equ:vdisp_model} and parameters  in \autoref{tab:coeff}.
\label{fig:sigma_ratio}}
\end{figure}

\subsection{The shape of the velocity ellipsoid}
Our best fit relations (see \autoref{equ:vdisp_model} and \autoref{tab:coeff}) can be used to estimate the ratio  $\sigma_z/\sigma_R$. These relations
suggest that there is a strong dependence of the ratio $\sigma_z/\sigma_R$ on age, metallicity, angular momentum, and height above the plane and this is shown in \autoref{fig:sigma_ratio}.
More precisely,
for 10 Gyr-old stars that are
in the plane and have solar metallicity, the ratio first decreases with guiding radius  to a minimum of 0.53 at $R_g=7.25$ kpc, and then increases with $R_g$. The ratio is greater than 0.7 for $R_g>12.0$ kpc. Also, the ratio increases monotonically with decrease in metallicity and increase of height.
If the vertical velocity dispersion is governed by scattering from GMCs, an equilibrium ratio of 0.62 is predicted, which can be attained by relatively old stars that had enough time to scatter. Simulations by \citet{2016MNRAS.462.1697A}
also suggest that spiral perturbations and GMCs can only
lead to ratios $\sigma_z/\sigma_R$ in range 0.5 to 0.7.
But the fact that we find $\sigma_z/\sigma_R$ to be greater than 0.7 in certain regions of the disc indicates that  processes other than spiral structure and GMCs could be affecting the vertical dispersion of stars there.

\subsection{Dependence of dispersions on metallicity}
We see that velocity dispersions increase with decreasing metallicity for any given age and angular momentum (\autoref{fig:sigma_vz_vr_fe_h_rg} and \autoref{fig:sigma_vz_vr_fe_h_age}).
Our current understanding of disc formation suggests that
the ISM probably had a negative metallicity gradient for a significant fraction of its lifetime \citep{2009MNRAS.396..203S,2018MNRAS.481.1645M}.
This suggests that at any given age  the metallicity should decrease with the birth radius of a star. Consequently,
velocity dispersions should increase with birth radius. This result is counter intuitive, as naively we expect the dispersion to decrease with any type of radius (\autoref{sec:dep_lz}).  Interestingly and importantly, we observe the dispersions to increase with birth radius for stars of all ages and angular momentum.  One reason for this could be
the conservation of vertical action, which will
happen for stars of any age or angular momentum.
\citet{2012MNRAS.422.1363S} demonstrated that vertical action is conserved for migrating stars. This conservation of vertical action leads to adiabatic heating/cooling of stars moving inwards/outwards \citep{2012A&A...548A.127M}. To demonstrate this,
let $E_z$ be the vertical energy, $\nu$
the vertical oscillation frequency, $\Sigma$ the surface density and $\sigma_z^2$ the vertical velocity dispersion.
Using some standard assumptions and approximations it is easy to show that
the vertical action $J_z=E_z/\nu=\sigma_z^2/\sqrt{2\pi G \Sigma}$ \citep[for details see \autoref{sec:disc_evol} and ][]{2012A&A...548A.127M}. When action is conserved, $\sigma_z \propto \Sigma^{1/4}$.
Hence, a stellar population born at radius $R_b$ with vertical velocity dispersion $\sigma_{zb}$ after migrating to orbits with guiding radius $R_g$ will end up with a dispersion $\sigma'_{zb}$ given by
\be
\sigma'_{zb}=\sigma_{zb} \exp\left(-(R_g-R_b)/4R_d\right), \label{equ:sigma_migration}
\ee
where $R_d$ is the scale length of surface density distribution.
From this expression it is easy to see that outward migration leads to cooling and inward migration leads to heating.

The question we are interested in is, for stars with a given angular momentum and age, is the dispersion of migrated stars higher or lower than that of non-migrated stars? The answer to this will depend on how $\sigma_{zb}$ varies with $R_b$ in \autoref{equ:sigma_migration}.
For simplicity let us assume that $\sigma_{zb}$ varies with $R_b$ just as
$\sigma_{z}$ varies with $R_g$, i.e., \autoref{equ:vdisp_model}.
Given that we find that the dispersion
is flat or rising for $R_g>R_{\odot}$, the inward migrators with $R_b>R_{\odot}$
are expected to be hotter than non-migrators. The outward migrators are also
expected to be hotter because $\sigma_{zb}$ increases with decreasing $R_b$ with a scale length (about $2 R_d$ see \autoref{sec:dep_lz}) that is smaller than $4R_d$. But it was demonstrated by \citet{2014ApJ...794..173V}, using an idealized simulation of a galaxy with spiral arms, that migrating stars are preferentially of low vertical dispersion, given that they spend more time in the plane \citep[see also][]{2018MNRAS.476.1561D}. This bias was also shown to be present for discs in cosmological simulations \citep{2016MNRAS.459..199G}. The bias makes it possible for the outward migrators to be cooler compared to non-migrators.
Other processes can also lead to an increase of dispersions
with birth radius. For example, effects like, disc satellite interaction,
infall of misalignment gas or reorientation of the disc axis, and warping (that as discussed earlier lead to flaring in the disc), all predict the dispersions to increase with birth radius.

\citet{2018MNRAS.481.1645M} had also reported an increase of
vertical dispersion with birth radius.
However, they
found the effect to be most prominent for older stars.
The slope of variation of dispersion with radius was found to
be positive for stars older than 8 Gyr and then flatten to zero at 6 Gyr and eventually turn negative for stars younger than 4 Gyr.
In contrast, we find the slope to be positive for all ages,
and additionally we also find a positive slope for the radial velocity dispersion, which they did not study.  \citet{2019MNRAS.489..176M} using mono age and mono metallicity populations reported an increase of
vertical dispersion with mean orbital radius for low-[$\alpha$/Fe] stars.  For the radial dispersion they reported an increase with mean orbital radius only for stars younger than 4 Gyr.
Given that mean orbital radius decreases monotonically with the metallicity for their populations, this means that their results can also be interpreted as an increase of dispersion with birth radius. In that case, for stars younger than 4 Gyr the slopes are opposite of that of \citet{2018MNRAS.481.1645M}.

Importantly, both \citet{2018MNRAS.481.1645M} and \citet{2019MNRAS.489..176M} had not factored out the dependence on angular momentum, which can have the
opposite effect, because for stars with angular momentum less than solar angular momentum, the dispersion increases with decrease of angular momentum. This could be responsible for the
differences between the above studies and differences with the results presented here.

If [$\alpha$/Fe] abundance is assumed to be
a good proxy for age, then our metallicity (or birth radius) trends can also be considered to be consistent with the findings of \citet{2020MNRAS.493.2952H}.
They studied the velocity dispersion as a function of [$\alpha$/Fe] abundance in different [Fe/H] bins using GALAH-DR2 data.
They found that the vertical dispersion increases with decrease of metallicity for any given
[$\alpha$/Fe] just as we find the same effect for any given age.

\subsection{Dependence of dispersions on height}
We find that velocity dispersions increase with height
for all angular momentum (\autoref{fig:sigma_vz_vr_pzgc_abs_rg}), ages (\autoref{fig:sigma_vz_vr_pzgc_abs_age})  and metallicities (\autoref{fig:sigma_vz_vr_pzgc_abs_fe_h}).
A positive slope is present for all
ages but it is much steeper for stars younger than
4 Gyr (\autoref{fig:sigma_vz_vr_pzgc_abs_age}).
Also the slope for  $\sigma_z$ is higher than that for $\sigma_R$
by about a factor of 2. This suggests that the ratio of $\sigma_z/\sigma_R$ also increases with height.  A non-zero slope
implies that the populations defined by
a specific age and metallicity are non-isothermal.
\citet{2019MNRAS.489..176M} also report a positive slope
for low $[\alpha/{\rm Fe}]$ populations, which
was found to flatten with age. For high-$[\alpha/{\rm Fe}]$ populations the slope was found to be zero, whereas we
find the high-$[\alpha/{\rm Fe}]$ stars to  have a positive slope.
As suggested by \citet{2019MNRAS.489..176M}, the non-isothermality could be related to the relatively large time scale for GMC heating as compared to the relatively
fast in-plane heating by spiral arms.
This is something that can be easily tested in idealized simulations by \citet{2016MNRAS.462.1697A}. However, as shown in \citet{1988A&A...192..117V}, isothermality is not necessary for constructing an equilibrium distribution. They show that for a self gravitating disc whose vertical density distribution is exponential, the vertical dispersion is found to increase with $|z|$.

\subsection{ Dependence of dispersions on $[\alpha/{\rm Fe}]$ abundance}
\begin{figure*}[tb]
\centering \includegraphics[width=0.99\textwidth]{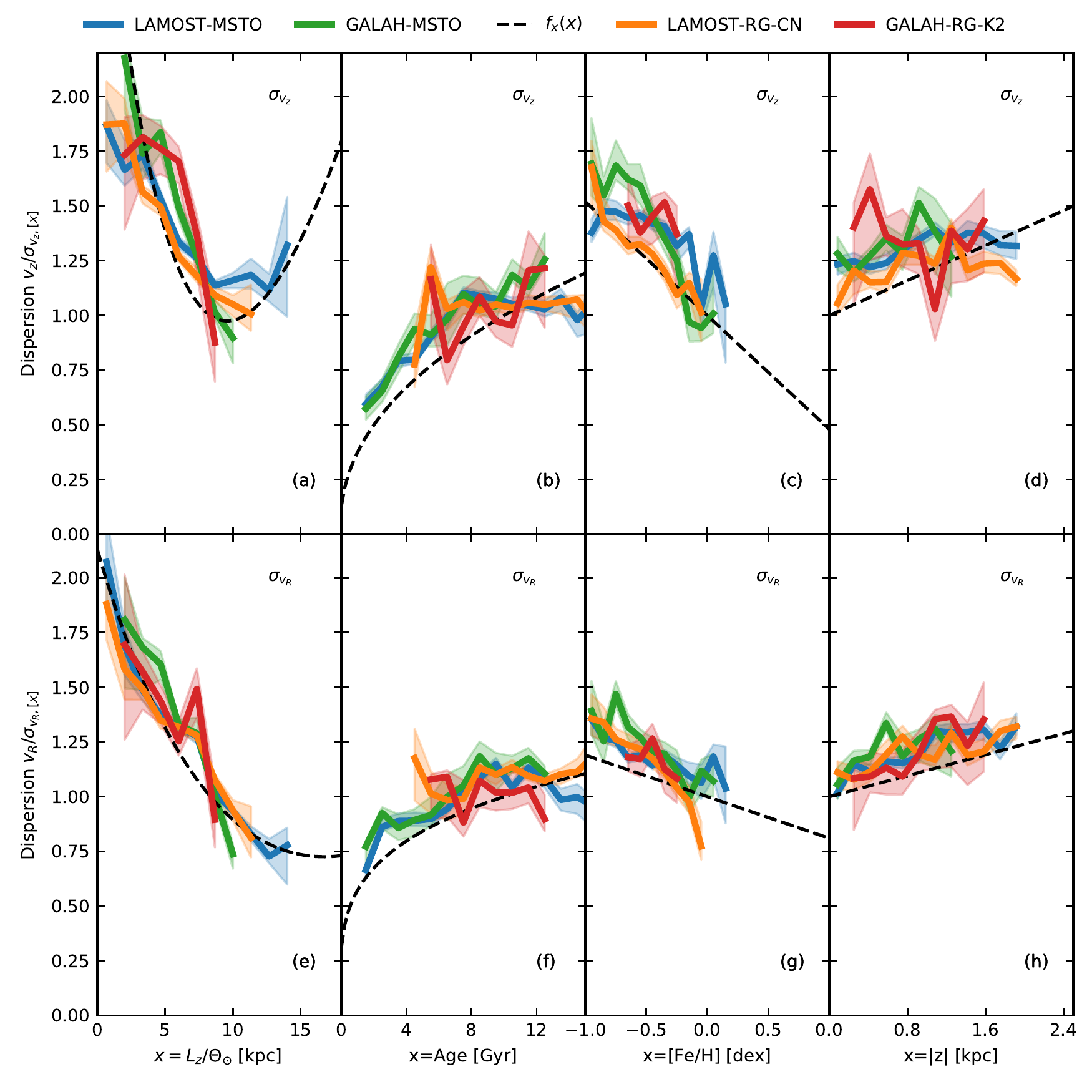}
\caption{Same as \autoref{fig:sigma_vz_vr} but for  stars with [$\alpha$/Fe]>0.25. The high-[$\alpha$/Fe] stars seem to follow the same relationship as low--[$\alpha$/Fe] stars, except for a small shift in the overall normalization.
\label{fig:sigma_vz_vr_al0.25}}
\end{figure*}

Contrary to claims by \citet{2019MNRAS.489..176M} that the velocity dispersion properties of the high-$[\alpha/{\rm Fe}]$ population is different from that of the low-$[\alpha/{\rm Fe}]$, we find very little difference between the two populations. We demonstrate this in \autoref{fig:sigma_vz_vr_al0.25}, where we plot the
dispersion as a function of various different independent variables.
The dashed lines show the best fit relation obeyed by all stars, which
can be considered as the relationship obeyed by low-[$\alpha$/Fe] stars, as
the sample of all stars is dominated by them. The high-$[\alpha/{\rm Fe}]$
stars are found to closely follow the dashed lines.
A constant shift of about 10\% can be seen between the
dashed and colored lines, indicating that the overall normalization may be
slightly different, but the profile shapes are very similar.
For old stars, the high-$[\alpha/{\rm Fe}]$ AVR seems to flatten more strongly
than for the low-$[\alpha/{\rm Fe}]$ AVR. We note that this effect is only prominent
for samples other than GALAH-MSTO; which also happen to have larger
age uncertainties as compared to the GALAH-MSTO sample.
Given that high-[$\alpha$/Fe] stars are expected
to lie in a narrow range in age, large uncertainties in age can
easily flatten the AVR. Hence, large age uncertainties seem to be
the most likely reason for the apparent flattening of the AVR.

\subsection{Testing the accuracy of the asteroseismic ages}
Giants are intrinsically bright and
hence for a given apparent magnitude limit they can probe a much larger Galactic volume
as compared to MSTO stars.
However, it is difficult to
estimate the ages of giants  from purely spectroscopic parameters.
Over the past decades, asteroseismology has attempted to break this barrier,
backed by very precise time series photometry from space missions like CoRot, {\it Kepler}, and K2.
However, measuring ages of a large numbers of stars often requires the use of asteroseismic scaling relations,
which are empirical. Testing the accuracy of these scaling relations
is complicated by the fact that it is difficult to get independent and precise measurements of mass or age of giants. A few techniques that have been
used to verify the asteroseismic ages are,
to assume that metal poor stars ([Fe/H]$<-1$) are older than 10 Gyr \citep{2014ApJ...785L..28E}, to use eclipsing binaries for estimating the mass \citep{2013ApJ...767...82G,2018MNRAS.476.3729B}, or to use cluster members to
estimate the age \citet{2012A&A...543A.106B}. While these studies suggest that the asteroseismic scaling relations overestimate masses by about 10\%, they are severely hampered by small number statistics.

An indirect means to verify asteroseismic ages is to rely on ensemble statistics --
for example, by comparing the mass distribution of stars against predictions
of population-synthesis-based models of the Galaxy.
\citet{2016ApJ...822...15S} and \citet{2017ApJ...835..163S} using {\it Kepler} data suggested that the
asteroseismic scaling-based masses were overestimated by about 10\% compared to model predictions.
However, a followup study by \citet{2019MNRAS.490.5335S}
using data from both {\it Kepler} and K2 showed that much of the tension between
observations and predictions is reduced after updating the metallicity of the thick disc to recent iron abundance measurements, and additionally taking $\alpha$-element abundance into account. However, we still do not have a Galactic model with all its free parameters tightly constrained. In fact certain parameters are degenerate. Hence, it is
useful to look for alternative methods to verify the asteroseismic ages.

Here we provide another method to verify asteroseismic ages
based on ensemble statistics --
comparing the velocity dispersion of stars conditional on age, metallicity, angular momentum and distance from the plane.
The underlying principle is that the velocity dispersion is a global Galactic property. Hence, groups of stars having same age, metallicity, angular momentum
and distance from the plane, should have same velocity dispersion,
irrespective of their stellar type, target selection and age estimation technique.
Using the above method, we find that the conditional velocity dispersion of the GALAH-RG-K2 asteroseismic sample is in agreement with that of the GALAH-MSTO sample.
The APOGEE-RG-KEPLER sample was found to have an offset with respect to
the GALAH-MSTO sample. However, part of this offest is due to APOGEE iron abundance
being higher by about 0.1 dex than GALAH. To account for the rest of the observed offset, the
APOGEE-RG-KEPLER asteroseismic ages had to be increased by about 10\%.
This bias is consistent, both in direction and amount, with earlier analysis
that compared the mass distribution of the {\it Kepler} sample with predictions from
stellar-population-synthesis based models \citep{2019MNRAS.490.5335S}.

We now demonstrate that a 10\% systematic in age
can easily stem from inaccuracies in measurement
of average seismic parameters $\Delta \nu$ or $\nu_{\rm max}$.
\citet{2018ApJS..239...32P} had shown that different methods for measuring
$\nu_{\rm max}$ and $\Delta{\nu}$ can have systematics of up to a few percent.
Given that the K2 light curve is much shorter (3 months as compared to 4 years) and is more noisy, we can also expect biases of upto a few percent in the seismic parameters
estimated from them. We show below that even a 1\% change in either $\Delta \nu$ or $\nu_{\rm max}$ can lead to a change of 10\% in age.
The age of a red giant star is primarily determined by the time it spends on the main sequence and is roughly $\tau_{\rm MS} \propto M/L(M) \propto M^{-3.8}$
for stars with $M<2{\rm M}_{\odot}$ \citep{1998gaas.book.....B}.
According to asteroseismic scaling relations $M \propto \nu_{\rm max}^3/\Delta \nu^{4}$, implying that the age depends on $\nu_{\rm max}$ and $\Delta \nu$
with a power greater than 10.

\section{Summary and Conclusions}
We have explored the fundamental relations governing the
radial and the vertical velocity dispersions of stars
in the Milky Way and discussed the dynamical processes that
might be responsible for them.
For the first time,
we present the joint dependence of the vertical and radial velocity dispersions on
age, angular momentum, metallicity and distance from the plane.
We compare and contrast results from three
different spectroscopic surveys, (GALAH, LAMOST, and APOGEE)
and three different stellar types (MSTO, asteroseismic giant, and
RGB stars).

Vertical and radial velocity dispersions depend upon at least 4 independent variables,
and these are age, angular momentum, metallicity and distance from the plane.
The joint dependence is well approximated
by a separable functional form that is a product of univariate functions, with each function
corresponding to one independent variable.
In other words, the dependence of the dispersions on each independent
variable is almost independent of the other variables.

The velocity dispersions increase with age following a power law, with exponent
$\beta_z=0.441 \pm 0.007$ for $\sigma_z$ and $\beta_R=0.251 \pm 0.006$ for $\sigma_R$.
These exponents are in good agreement with idealized simulations of
\citet{2016MNRAS.459.3326A} where the disc heating is due to scattering
by bar, spiral arms and GMCs.

The velocity dispersions show a non-monotonic behaviour with $L_z$.
They decrease with $L_z$ until about solar angular momentum, thereafter,
$\sigma_R$ flattens, while $\sigma_z$ increases. The flattening
at large $L_z$ could be due to a non-zero floor on the
intrinsic birth dispersion of stars. However,
the cause for the rise of $\sigma_z$ at large $L_z$ is
not fully understood. Idealized simulations having a bar,
spiral structure and GMCs do not show such an effect.
However, cosmological simulations by \citet{2016MNRAS.459..199G}
do show such an effect, where the heating is
primarily attributed to a bar and orbiting satellites.
However, other factors, like
warps, the infall of misaligned gas and reorientation of the disc,
can also be responsible for the above effect.

The velocity dispersions decrease almost linearly with metallicity, or in
other words the velocity dispersion increases with birth radius.
We show that this can be explained
by the conservation of vertical
action in stars undergoing radial migration.
However, for this to work, stars migrating
outwards from the inner regions
should be preferentially of low velocity dispersion -- the
so called "provenance bias" as  discussed by \citet{2014ApJ...794..173V}
and also reported by others \citep{2016MNRAS.459..199G,  2018MNRAS.476.1561D}.

The velocity dispersions increase almost linearly with distance from the
plane. This effect is more prominent for younger stars.
Additionally, the effect is stronger for $\sigma_z$ than for
$\sigma_R$. This agrees with findings of \citet{2019MNRAS.489..176M} using
APOGEE giants. Spiral arms are responsible for in-plane scattering,
while GMCs are though to redirect the planar motion into the vertical direction \citep{1990MNRAS.245..305J}.
As suggested by \citet{2019MNRAS.489..176M},
the longer time scale associated with GMC heating as compared to
spiral heating could be responsible for the observed non-isothermality.

A particularly useful aspect of identifying the set of independent variables
that govern the velocity dispersion, is that if the dispersion is characterized in terms of these variables then it is almost independent of the target selection function. This provides a means to not only compare results from different observational data sets (e.g  to test systematics in spectroscopic stellar parameters between different surveys, or systematics  between different age estimation techniques), but also to compare observational results with theoretical predictions.
We take advantage of this fact to show that GALAH and LAMOST
results are in agreement with each other and that results from different stellar types
(MSTO and giant stars)
are also in agreement with each other.
The ages of giant stars have been estimated using the asteroseismic scaling
relations either directly (for K2 stars) or indirectly (for LAMOST RGBs).
It is difficult to verify the asteroseismic scaling relations
due to a shortage of independent estimates of stellar mass or age.
In this sense, we provide a new
technique based on ensemble statistics
to verify the accuracy of the asteroseismic ages.

The velocity dispersion of the APOGEE data set of asteroseismic giants from {\it Kepler}
was found to be systematically different from our derived relations. We identify two possible reasons for this.
First, the metallicity of APOGEE
giants is systematically lower by about 0.1 dex with respect GALAH and LAMOST. Second, it is possible that the average asteroseismic parameters derived from {\it Kepler} data have some systematics with respect to those derived from the K2 data, given that
the light curves from K2 are much shorter (3 months as compared to 4 years) and  noisier.

Finally, we find that all stars, irrespective of them being old
or having high-[$\alpha$/Fe],
follow the same relations for velocity dispersion. In other words,
no special provision is needed to accommodate the
thick disc stars. The AVR of stars in the solar neighborhood
does show a break from a pure power law for stars older
than 8 Gyr. However, when the angular momentum and metallicity
of these stars is taken into account no such break is seen.
The apparent break
is due to older stars having systematically lower angular momentum.

\acknowledgments
SS is funded by a Senior Fellowship (University of Sydney), an ASTRO-3D Research Fellowship and JBH's Laureate Fellowship from the Australian Research Council (ARC).
JBH's research team is supported by an ARC Laureate Fellowship (FL140100278) and funds from ASTRO-3D. MJH is supported by an ASTRO-3D 4-year Research Fellowship.
DS is the recipient of an ARC Future Fellowship (project number FT1400147).
SB and KL acknowledge funds from the Alexander von Humboldt Foundation in the framework of the Sofja Kovalevskaja Award endowed by the Federal Ministry of Education and Research.
KL acknowledges funds from the Swedish Research Council (Grant 2015-00415\_3) and Marie Sklodowska Curie Actions (Cofund Project INCA 600398).
JK, KC and TZ acknowledges financial support from the Slovenian Research Agency (research core funding No. P1-0188).  DMN was supported by the Allan C. and Dorothy H. Davis Fellowship. JZ acknowledges support from NASA grants 80NSSC18K0391 and NNX17AJ40G.
DH acknowledges support from the Alfred P. Sloan Foundation and the National Aeronautics and Space Administration (80NSSC19K0108).
The GALAH Survey is supported by the ARC Centre of Excellence for All Sky Astrophysics in 3 Dimensions (ASTRO 3D), through project CE170100013.
This work has made use of data acquired through the Australian Astronomical Observatory, under programs: GALAH, TESS-HERMES and K2-HERMES. We acknowledge the traditional owners of the land on which the AAT stands, the Gamilaraay people, and pay our respects to elders past and present.

This work has made use of data from the European Space Agency (ESA) mission
{\it Gaia} (\url{https://www.cosmos.esa.int/gaia}), processed by the {\it Gaia}
Data Processing and Analysis Consortium (DPAC,
\url{https://www.cosmos.esa.int/web/gaia/dpac/consortium}). Funding for the DPAC
has been provided by national institutions, in particular the institutions
participating in the {\it Gaia} Multilateral Agreement.

This work has made use of data from SDSS-III.
Funding for SDSS-III has been provided by the Alfred P. Sloan Foundation, the Participating Institutions, the National Science Foundation, and the U.S. Department of Energy Office of Science. The SDSS-III web site is http://www.sdss3.org/.

This work has made use of Guoshoujing Telescope (the Large Sky Area Multi-Object Fiber Spectroscopic Telescope LAMOST) which is a National Major Scientific Project built by the Chinese Academy of Sciences. Funding for the project has been provided by the National Development and Reform Commission. LAMOST is operated and managed by the National Astronomical Observatories, Chinese Academy of Sciences.

\bibliographystyle{yahapj}
\bibliography{references}

\begin{thebibliography}{}
\providecommand\natexlab[1]{#1}
\providecommand\JournalTitle[1]{#1}

\bibitem[{{Aumer} {et~al.}(2016{\natexlab{a}}){Aumer}, {Binney}, \&
  {Sch{\"o}nrich}}]{2016MNRAS.462.1697A}
{Aumer}, M., {Binney}, J., \& {Sch{\"o}nrich}, R. 2016{\natexlab{a}},
  \href{http://dx.doi.org/10.1093/mnras/stw1639}{\JournalTitle{\mnras}, 462,
  1697}

\bibitem[{{Aumer} {et~al.}(2016{\natexlab{b}}){Aumer}, {Binney}, \&
  {Sch{\"o}nrich}}]{2016MNRAS.459.3326A}
---. 2016{\natexlab{b}},
  \href{http://dx.doi.org/10.1093/mnras/stw777}{\JournalTitle{\mnras}, 459,
  3326}

\bibitem[{{Aumer} \& {Binney}(2009)}]{2009MNRAS.397.1286A}
{Aumer}, M., \& {Binney}, J.~J. 2009,
  \href{http://dx.doi.org/10.1111/j.1365-2966.2009.15053.x}{\JournalTitle{\mnras},
  397, 1286}

\bibitem[{{Aumer} \& {White}(2013)}]{2013MNRAS.428.1055A}
{Aumer}, M., \& {White}, S. D.~M. 2013,
  \href{http://dx.doi.org/10.1093/mnras/sts083}{\JournalTitle{\mnras}, 428,
  1055}

\bibitem[{{Aumer} {et~al.}(2013){Aumer}, {White}, {Naab}, \&
  {Scannapieco}}]{2013MNRAS.434.3142A}
{Aumer}, M., {White}, S. D.~M., {Naab}, T., \& {Scannapieco}, C. 2013,
  \href{http://dx.doi.org/10.1093/mnras/stt1230}{\JournalTitle{\mnras}, 434,
  3142}

\bibitem[{{Barbanis} \& {Woltjer}(1967)}]{1967ApJ...150..461B}
{Barbanis}, B., \& {Woltjer}, L. 1967,
  \href{http://dx.doi.org/10.1086/149349}{\JournalTitle{\apj}, 150, 461}

\bibitem[{{Binney}(2012)}]{2012MNRAS.426.1328B}
{Binney}, J. 2012,
  \href{http://dx.doi.org/10.1111/j.1365-2966.2012.21692.x}{\JournalTitle{\mnras},
  426, 1328}

\bibitem[{{Binney} \& {Merrifield}(1998)}]{1998gaas.book.....B}
{Binney}, J., \& {Merrifield}, M. 1998, {Galactic Astronomy} (Princeton
  University Press)

\bibitem[{{Binney} \& {Tremaine}(2008)}]{2008gady.book.....B}
{Binney}, J., \& {Tremaine}, S. 2008, {Galactic Dynamics: Second Edition}
  (Princeton Series in Astrophysics)

\bibitem[{{Bland-Hawthorn} {et~al.}(2010){Bland-Hawthorn}, {Krumholz}, \&
  {Freeman}}]{2010ApJ...713..166B}
{Bland-Hawthorn}, J., {Krumholz}, M.~R., \& {Freeman}, K. 2010,
  \href{http://dx.doi.org/10.1088/0004-637X/713/1/166}{\JournalTitle{\apj},
  713, 166}

\bibitem[{{Bland-Hawthorn} {et~al.}(2019){Bland-Hawthorn}, {Sharma},
  {Tepper-Garcia}, {Binney}, {Freeman}, {Hayden}, {Kos}, {De Silva}, {Ellis},
  {Lewis}, {Asplund}, {Buder}, {Casey}, {D'Orazi}, {Duong}, {Khanna}, {Lin},
  {Lind}, {Martell}, {Ness}, {Simpson}, {Zucker}, {Zwitter}, {Kafle},
  {Quillen}, {Ting}, \& {Wyse}}]{2019MNRAS.486.1167B}
{Bland-Hawthorn}, J., {Sharma}, S., {Tepper-Garcia}, T., {et~al.} 2019,
  \href{http://dx.doi.org/10.1093/mnras/stz217}{\JournalTitle{\mnras}, 486,
  1167}

\bibitem[{{Bournaud} {et~al.}(2009){Bournaud}, {Elmegreen}, \&
  {Martig}}]{2009ApJ...707L...1B}
{Bournaud}, F., {Elmegreen}, B.~G., \& {Martig}, M. 2009,
  \href{http://dx.doi.org/10.1088/0004-637X/707/1/L1}{\JournalTitle{\apjl},
  707, L1}

\bibitem[{{Brogaard} {et~al.}(2012){Brogaard}, {VandenBerg}, {Bruntt},
  {Grundahl}, {Frandsen}, {Bedin}, {Milone}, {Dotter}, {Feiden}, {Stetson},
  {Sandquist}, {Miglio}, {Stello}, \& {Jessen-Hansen}}]{2012A&A...543A.106B}
{Brogaard}, K., {VandenBerg}, D.~A., {Bruntt}, H., {et~al.} 2012,
  \href{http://dx.doi.org/10.1051/0004-6361/201219196}{\JournalTitle{\aap},
  543, A106}

\bibitem[{{Brogaard} {et~al.}(2018){Brogaard}, {Hansen}, {Miglio}, {Slumstrup},
  {Frandsen}, {Jessen-Hansen}, {Lund}, {Bossini}, {Thygesen}, {Davies},
  {Chaplin}, {Arentoft}, {Bruntt}, {Grundahl}, \&
  {Handberg}}]{2018MNRAS.476.3729B}
{Brogaard}, K., {Hansen}, C.~J., {Miglio}, A., {et~al.} 2018,
  \href{http://dx.doi.org/10.1093/mnras/sty268}{\JournalTitle{\mnras}, 476,
  3729}

\bibitem[{{Buder} {et~al.}(2018){Buder}, {Asplund}, {Duong}, {Kos}, {Lind},
  {Ness}, {Sharma}, {Bland -Hawthorn}, {Casey}, {de Silva}, {D'Orazi},
  {Freeman}, {Lewis}, {Lin}, {Martell}, {Schlesinger}, {Simpson}, {Zucker},
  {Zwitter}, {Amarsi}, {Anguiano}, {Carollo}, {Casagrande}, {{\v{C}}otar},
  {Cottrell}, {da Costa}, {Gao}, {Hayden}, {Horner}, {Ireland}, {Kafle},
  {Munari}, {Nataf}, {Nordlander}, {Stello}, {Ting}, {Traven}, {Watson},
  {Wittenmyer}, {Wyse}, {Yong}, {Zinn}, {{\v{Z}}erjal}, \& {Galah
  Collaboration}}]{2018MNRAS.478.4513B}
{Buder}, S., {Asplund}, M., {Duong}, L., {et~al.} 2018,
  \href{http://dx.doi.org/10.1093/mnras/sty1281}{\JournalTitle{\mnras}, 478,
  4513}

\bibitem[{{Carlberg} \& {Sellwood}(1985)}]{1985ApJ...292...79C}
{Carlberg}, R.~G., \& {Sellwood}, J.~A. 1985,
  \href{http://dx.doi.org/10.1086/163134}{\JournalTitle{\apj}, 292, 79}

\bibitem[{{Daniel} \& {Wyse}(2018)}]{2018MNRAS.476.1561D}
{Daniel}, K.~J., \& {Wyse}, R. F.~G. 2018,
  \href{http://dx.doi.org/10.1093/mnras/sty199}{\JournalTitle{\mnras}, 476,
  1561}

\bibitem[{{De Silva} {et~al.}(2015){De Silva}, {Freeman}, {Bland-Hawthorn},
  {Martell}, {de Boer}, {Asplund}, {Keller}, {Sharma}, {Zucker}, {Zwitter},
  {Anguiano}, {Bacigalupo}, {Bayliss}, {Beavis}, {Bergemann}, {Campbell},
  {Cannon}, {Carollo}, {Casagrande}, {Casey}, {Da Costa}, {D'Orazi}, {Dotter},
  {Duong}, {Heger}, {Ireland}, {Kafle}, {Kos}, {Lattanzio}, {Lewis}, {Lin},
  {Lind}, {Munari}, {Nataf}, {O'Toole}, {Parker}, {Reid}, {Schlesinger},
  {Sheinis}, {Simpson}, {Stello}, {Ting}, {Traven}, {Watson}, {Wittenmyer},
  {Yong}, \& {{\v{Z}}erjal}}]{2015MNRAS.449.2604D}
{De Silva}, G.~M., {Freeman}, K.~C., {Bland-Hawthorn}, J., {et~al.} 2015,
  \href{http://dx.doi.org/10.1093/mnras/stv327}{\JournalTitle{\mnras}, 449,
  2604}

\bibitem[{{De Simone} {et~al.}(2004){De Simone}, {Wu}, \&
  {Tremaine}}]{2004MNRAS.350..627D}
{De Simone}, R., {Wu}, X., \& {Tremaine}, S. 2004,
  \href{http://dx.doi.org/10.1111/j.1365-2966.2004.07675.x}{\JournalTitle{\mnras},
  350, 627}

\bibitem[{{Deng} {et~al.}(2012){Deng}, {Newberg}, {Liu}, {Carlin}, {Beers},
  {Chen}, {Chen}, {Christlieb}, {Grillmair}, {Guhathakurta}, {Han}, {Hou},
  {Lee}, {L{\'e}pine}, {Li}, {Liu}, {Pan}, {Sellwood}, {Wang}, {Wang}, {Yang},
  {Yanny}, {Zhang}, {Zhang}, {Zheng}, \& {Zhu}}]{2012RAA....12..735D}
{Deng}, L.-C., {Newberg}, H.~J., {Liu}, C., {et~al.} 2012,
  \href{http://dx.doi.org/10.1088/1674-4527/12/7/003}{\JournalTitle{Research in
  Astronomy and Astrophysics}, 12, 735}

\bibitem[{{Edvardsson} {et~al.}(1993){Edvardsson}, {Andersen}, {Gustafsson},
  {Lambert}, {Nissen}, \& {Tomkin}}]{1993A&A...275..101E}
{Edvardsson}, B., {Andersen}, J., {Gustafsson}, B., {et~al.} 1993,
  \JournalTitle{\aap}, 500, 391

\bibitem[{{Epstein} {et~al.}(2014){Epstein}, {Elsworth}, {Johnson}, {Shetrone},
  {Mosser}, {Hekker}, {Tayar}, {Harding}, {Pinsonneault}, {Silva Aguirre},
  {Basu}, {Beers}, {Bizyaev}, {Bedding}, {Chaplin}, {Frinchaboy},
  {Garc{\'\i}a}, {Garc{\'\i}a P{\'e}rez}, {Hearty}, {Huber}, {Ivans},
  {Majewski}, {Mathur}, {Nidever}, {Serenelli}, {Schiavon}, {Schneider},
  {Sch{\"o}nrich}, {Sobeck}, {Stassun}, {Stello}, \&
  {Zasowski}}]{2014ApJ...785L..28E}
{Epstein}, C.~R., {Elsworth}, Y.~P., {Johnson}, J.~A., {et~al.} 2014,
  \href{http://dx.doi.org/10.1088/2041-8205/785/2/L28}{\JournalTitle{\apjl},
  785, L28}

\bibitem[{{F{\"o}rster Schreiber} {et~al.}(2009){F{\"o}rster Schreiber},
  {Genzel}, {Bouch{\'e}}, {Cresci}, {Davies}, {Buschkamp}, {Shapiro},
  {Tacconi}, {Hicks}, {Genel}, {Shapley}, {Erb}, {Steidel}, {Lutz},
  {Eisenhauer}, {Gillessen}, {Sternberg}, {Renzini}, {Cimatti}, {Daddi},
  {Kurk}, {Lilly}, {Kong}, {Lehnert}, {Nesvadba}, {Verma}, {McCracken},
  {Arimoto}, {Mignoli}, \& {Onodera}}]{2009ApJ...706.1364F}
{F{\"o}rster Schreiber}, N.~M., {Genzel}, R., {Bouch{\'e}}, N., {et~al.} 2009,
  \href{http://dx.doi.org/10.1088/0004-637X/706/2/1364}{\JournalTitle{\apj},
  706, 1364}

\bibitem[{{Frankel} {et~al.}(2019){Frankel}, {Sanders}, {Rix}, {Ting}, \&
  {Ness}}]{2019ApJ...884...99F}
{Frankel}, N., {Sanders}, J., {Rix}, H.-W., {Ting}, Y.-S., \& {Ness}, M. 2019,
  \href{http://dx.doi.org/10.3847/1538-4357/ab4254}{\JournalTitle{\apj}, 884,
  99}

\bibitem[{{Freeman} \& {Bland-Hawthorn}(2002)}]{2002ARA&A..40..487F}
{Freeman}, K., \& {Bland-Hawthorn}, J. 2002,
  \href{http://dx.doi.org/10.1146/annurev.astro.40.060401.093840}{\JournalTitle{\araa},
  40, 487}

\bibitem[{{Freeman}(1991)}]{1991dodg.conf...15F}
{Freeman}, K.~C. 1991, in Dynamics of Disc Galaxies, ed. B.~{Sundelius}, 15

\bibitem[{{Gaia Collaboration} {et~al.}(2018){Gaia Collaboration}, {Brown},
  {Vallenari}, {Prusti}, {de Bruijne}, {Babusiaux}, {Bailer-Jones}, {Biermann},
  {Evans}, {Eyer}, {Jansen}, {Jordi}, {Klioner}, {Lammers}, {Lindegren},
  {Luri}, {Mignard}, {Panem}, {Pourbaix}, {Randich}, {Sartoretti}, {Siddiqui},
  {Soubiran}, {van Leeuwen}, {Walton}, {Arenou}, {Bastian}, {Cropper},
  {Drimmel}, {Katz}, {Lattanzi}, {Bakker}, {Cacciari}, {Casta{\~n}eda},
  {Chaoul}, {Cheek}, {De Angeli}, {Fabricius}, {Guerra}, {Holl}, {Masana},
  {Messineo}, {Mowlavi}, {Nienartowicz}, {Panuzzo}, {Portell}, {Riello},
  {Seabroke}, {Tanga}, {Th{\'e}venin}, {Gracia-Abril}, {Comoretto},
  {Garcia-Reinaldos}, {Teyssier}, {Altmann}, {Andrae}, {Audard},
  {Bellas-Velidis}, {Benson}, {Berthier}, {Blomme}, {Burgess}, {Busso},
  {Carry}, {Cellino}, {Clementini}, {Clotet}, {Creevey}, {Davidson}, {De
  Ridder}, {Delchambre}, {Dell'Oro}, {Ducourant},
  {Fern{\'a}ndez-Hern{\'a}ndez}, {Fouesneau}, {Fr{\'e}mat}, {Galluccio},
  {Garc{\'\i}a-Torres}, {Gonz{\'a}lez-N{\'u}{\~n}ez}, {Gonz{\'a}lez-Vidal},
  {Gosset}, {Guy}, {Halbwachs}, {Hambly}, {Harrison}, {Hern{\'a}ndez},
  {Hestroffer}, {Hodgkin}, {Hutton}, {Jasniewicz}, {Jean-Antoine-Piccolo},
  {Jordan}, {Korn}, {Krone-Martins}, {Lanzafame}, {Lebzelter}, {L{\"o}ffler},
  {Manteiga}, {Marrese}, {Mart{\'\i}n-Fleitas}, {Moitinho}, {Mora}, {Muinonen},
  {Osinde}, {Pancino}, {Pauwels}, {Petit}, {Recio-Blanco}, {Richards},
  {Rimoldini}, {Robin}, {Sarro}, {Siopis}, {Smith}, {Sozzetti}, {S{\"u}veges},
  {Torra}, {van Reeven}, {Abbas}, {Abreu Aramburu}, {Accart}, {Aerts},
  {Altavilla}, {{\'A}lvarez}, {Alvarez}, {Alves}, {Anderson}, {Andrei},
  {Anglada Varela}, {Antiche}, {Antoja}, {Arcay}, {Astraatmadja}, {Bach},
  {Baker}, {Balaguer-N{\'u}{\~n}ez}, {Balm}, {Barache}, {Barata}, {Barbato},
  {Barblan}, {Barklem}, {Barrado}, {Barros}, {Barstow}, {Bartholom{\'e}
  Mu{\~n}oz}, {Bassilana}, {Becciani}, {Bellazzini}, {Berihuete}, {Bertone},
  {Bianchi}, {Bienaym{\'e}}, {Blanco-Cuaresma}, {Boch}, {Boeche}, {Bombrun},
  {Borrachero}, {Bossini}, {Bouquillon}, {Bourda}, {Bragaglia}, {Bramante},
  {Breddels}, {Bressan}, {Brouillet}, {Br{\"u}semeister}, {Brugaletta},
  {Bucciarelli}, {Burlacu}, {Busonero}, {Butkevich}, {Buzzi}, {Caffau},
  {Cancelliere}, {Cannizzaro}, {Cantat-Gaudin}, {Carballo}, {Carlucci},
  {Carrasco}, {Casamiquela}, {Castellani}, {Castro-Ginard}, {Charlot},
  {Chemin}, {Chiavassa}, {Cocozza}, {Costigan}, {Cowell}, {Crifo}, {Crosta},
  {Crowley}, {Cuypers}, {Dafonte}, {Damerdji}, {Dapergolas}, {David}, {David},
  {de Laverny}, {De Luise}, {De March}, {de Martino}, {de Souza}, {de Torres},
  {Debosscher}, {del Pozo}, {Delbo}, {Delgado}, {Delgado}, {Di Matteo},
  {Diakite}, {Diener}, {Distefano}, {Dolding}, {Drazinos}, {Dur{\'a}n},
  {Edvardsson}, {Enke}, {Eriksson}, {Esquej}, {Eynard Bontemps}, {Fabre},
  {Fabrizio}, {Faigler}, {Falc{\~a}o}, {Farr{\`a}s Casas}, {Federici},
  {Fedorets}, {Fernique}, {Figueras}, {Filippi}, {Findeisen}, {Fonti},
  {Fraile}, {Fraser}, {Fr{\'e}zouls}, {Gai}, {Galleti}, {Garabato},
  {Garc{\'\i}a-Sedano}, {Garofalo}, {Garralda}, {Gavel}, {Gavras}, {Gerssen},
  {Geyer}, {Giacobbe}, {Gilmore}, {Girona}, {Giuffrida}, {Glass}, {Gomes},
  {Granvik}, {Gueguen}, {Guerrier}, {Guiraud}, {Guti{\'e}rrez-S{\'a}nchez},
  {Haigron}, {Hatzidimitriou}, {Hauser}, {Haywood}, {Heiter}, {Helmi}, {Heu},
  {Hilger}, {Hobbs}, {Hofmann}, {Holland}, {Huckle}, {Hypki}, {Icardi},
  {Jan{\ss}en}, {Jevardat de Fombelle}, {Jonker}, {Juh{\'a}sz}, {Julbe},
  {Karampelas}, {Kewley}, {Klar}, {Kochoska}, {Kohley}, {Kolenberg},
  {Kontizas}, {Kontizas}, {Koposov}, {Kordopatis}, {Kostrzewa-Rutkowska},
  {Koubsky}, {Lambert}, {Lanza}, {Lasne}, {Lavigne}, {Le Fustec}, {Le
  Poncin-Lafitte}, {Lebreton}, {Leccia}, {Leclerc}, {Lecoeur-Taibi},
  {Lenhardt}, {Leroux}, {Liao}, {Licata}, {Lindstr{\o}m}, {Lister}, {Livanou},
  {Lobel}, {L{\'o}pez}, {Managau}, {Mann}, {Mantelet}, {Marchal}, {Marchant},
  {Marconi}, {Marinoni}, {Marschalk{\'o}}, {Marshall}, {Martino}, {Marton},
  {Mary}, {Massari}, {Matijevi{\v{c}}}, {Mazeh}, {McMillan}, {Messina},
  {Michalik}, {Millar}, {Molina}, {Molinaro}, {Moln{\'a}r}, {Montegriffo},
  {Mor}, {Morbidelli}, {Morel}, {Morris}, {Mulone}, {Muraveva}, {Musella},
  {Nelemans}, {Nicastro}, {Noval}, {O'Mullane}, {Ord{\'e}novic},
  {Ord{\'o}{\~n}ez-Blanco}, {Osborne}, {Pagani}, {Pagano}, {Pailler},
  {Palacin}, {Palaversa}, {Panahi}, {Pawlak}, {Piersimoni}, {Pineau}, {Plachy},
  {Plum}, {Poggio}, {Poujoulet}, {Pr{\v{s}}a}, {Pulone}, {Racero}, {Ragaini},
  {Rambaux}, {Ramos-Lerate}, {Regibo}, {Reyl{\'e}}, {Riclet}, {Ripepi}, {Riva},
  {Rivard}, {Rixon}, {Roegiers}, {Roelens}, {Romero-G{\'o}mez}, {Rowell},
  {Royer}, {Ruiz-Dern}, {Sadowski}, {Sagrist{\`a} Sell{\'e}s}, {Sahlmann},
  {Salgado}, {Salguero}, {Sanna}, {Santana-Ros}, {Sarasso}, {Savietto},
  {Schultheis}, {Sciacca}, {Segol}, {Segovia}, {S{\'e}gransan}, {Shih},
  {Siltala}, {Silva}, {Smart}, {Smith}, {Solano}, {Solitro}, {Sordo}, {Soria
  Nieto}, {Souchay}, {Spagna}, {Spoto}, {Stampa}, {Steele},
  {Steidelm{\"u}ller}, {Stephenson}, {Stoev}, {Suess}, {Surdej}, {Szabados},
  {Szegedi-Elek}, {Tapiador}, {Taris}, {Tauran}, {Taylor}, {Teixeira},
  {Terrett}, {Teyssand ier}, {Thuillot}, {Titarenko}, {Torra Clotet}, {Turon},
  {Ulla}, {Utrilla}, {Uzzi}, {Vaillant}, {Valentini}, {Valette}, {van Elteren},
  {Van Hemelryck}, {van Leeuwen}, {Vaschetto}, {Vecchiato}, {Veljanoski},
  {Viala}, {Vicente}, {Vogt}, {von Essen}, {Voss}, {Votruba}, {Voutsinas},
  {Walmsley}, {Weiler}, {Wertz}, {Wevers}, {Wyrzykowski}, {Yoldas},
  {{\v{Z}}erjal}, {Ziaeepour}, {Zorec}, {Zschocke}, {Zucker}, {Zurbach}, \&
  {Zwitter}}]{2018A&A...616A...1G}
{Gaia Collaboration}, {Brown}, A.~G.~A., {Vallenari}, A., {et~al.} 2018,
  \href{http://dx.doi.org/10.1051/0004-6361/201833051}{\JournalTitle{\aap},
  616, A1}

\bibitem[{{Gaulme} {et~al.}(2013){Gaulme}, {McKeever}, {Rawls}, {Jackiewicz},
  {Mosser}, \& {Guzik}}]{2013ApJ...767...82G}
{Gaulme}, P., {McKeever}, J., {Rawls}, M.~L., {et~al.} 2013,
  \href{http://dx.doi.org/10.1088/0004-637X/767/1/82}{\JournalTitle{\apj}, 767,
  82}

\bibitem[{{Grand} {et~al.}(2016){Grand}, {Springel}, {G{\'o}mez}, {Marinacci},
  {Pakmor}, {Campbell}, \& {Jenkins}}]{2016MNRAS.459..199G}
{Grand}, R. J.~J., {Springel}, V., {G{\'o}mez}, F.~A., {et~al.} 2016,
  \href{http://dx.doi.org/10.1093/mnras/stw601}{\JournalTitle{\mnras}, 459,
  199}

\bibitem[{{Grenon}(1972)}]{1972ade..coll...55G}
{Grenon}, M. 1972, in IAU Colloq. 17: Age des Etoiles, ed. G.~{Cayrel de
  Strobel} \& A.~M. {Delplace}, 55

\bibitem[{{H{\"a}nninen} \& {Flynn}(2002)}]{2002MNRAS.337..731H}
{H{\"a}nninen}, J., \& {Flynn}, C. 2002,
  \href{http://dx.doi.org/10.1046/j.1365-8711.2002.05956.x}{\JournalTitle{\mnras},
  337, 731}

\bibitem[{{Hayden} {et~al.}(2020){Hayden}, {Bland-Hawthorn}, {Sharma},
  {Freeman}, {Kos}, {Buder}, {Anguiano}, {Asplund}, {Chen}, {De Silva},
  {Khanna}, {Lin}, {Horner}, {Martell}, {Ting}, {Wyse}, {Zucker}, \&
  {Zwitter}}]{2020MNRAS.493.2952H}
{Hayden}, M.~R., {Bland-Hawthorn}, J., {Sharma}, S., {et~al.} 2020,
  \href{http://dx.doi.org/10.1093/mnras/staa335}{\JournalTitle{\mnras}, 493,
  2952}

\bibitem[{{Ida} {et~al.}(1993){Ida}, {Kokubo}, \&
  {Makino}}]{1993MNRAS.263..875I}
{Ida}, S., {Kokubo}, E., \& {Makino}, J. 1993,
  \href{http://dx.doi.org/10.1093/mnras/263.4.875}{\JournalTitle{\mnras}, 263,
  875}

\bibitem[{{Jenkins} \& {Binney}(1990)}]{1990MNRAS.245..305J}
{Jenkins}, A., \& {Binney}, J. 1990, \JournalTitle{\mnras}, 245, 305

\bibitem[{{Juri{\'c}} {et~al.}(2008){Juri{\'c}}, {Ivezi{\'c}}, {Brooks},
  {Lupton}, {Schlegel}, {Finkbeiner}, {Padmanabhan}, {Bond}, {Sesar},
  {Rockosi}, {Knapp}, {Gunn}, {Sumi}, {Schneider}, {Barentine}, {Brewington},
  {Brinkmann}, {Fukugita}, {Harvanek}, {Kleinman}, {Krzesinski}, {Long},
  {Neilsen}, {Nitta}, {Snedden}, \& {York}}]{2008ApJ...673..864J}
{Juri{\'c}}, M., {Ivezi{\'c}}, {\v{Z}}., {Brooks}, A., {et~al.} 2008,
  \href{http://dx.doi.org/10.1086/523619}{\JournalTitle{\apj}, 673, 864}

\bibitem[{{Kallinger} {et~al.}(2010){Kallinger}, {Mosser}, {Hekker}, {Huber},
  {Stello}, {Mathur}, {Basu}, {Bedding}, {Chaplin}, {De Ridder}, {Elsworth},
  {Frand sen}, {Garc{\'\i}a}, {Gruberbauer}, {Matthews}, {Borucki}, {Bruntt},
  {Christensen-Dalsgaard}, {Gilliland}, {Kjeldsen}, \&
  {Koch}}]{2010A&A...522A...1K}
{Kallinger}, T., {Mosser}, B., {Hekker}, S., {et~al.} 2010,
  \href{http://dx.doi.org/10.1051/0004-6361/201015263}{\JournalTitle{\aap},
  522, A1}

\bibitem[{{Kallinger} {et~al.}(2014){Kallinger}, {De Ridder}, {Hekker},
  {Mathur}, {Mosser}, {Gruberbauer}, {Garc{\'\i}a}, {Karoff}, \&
  {Ballot}}]{2014A&A...570A..41K}
{Kallinger}, T., {De Ridder}, J., {Hekker}, S., {et~al.} 2014,
  \href{http://dx.doi.org/10.1051/0004-6361/201424313}{\JournalTitle{\aap},
  570, A41}

\bibitem[{{Kazantzidis} {et~al.}(2008){Kazantzidis}, {Bullock}, {Zentner},
  {Kravtsov}, \& {Moustakas}}]{2008ApJ...688..254K}
{Kazantzidis}, S., {Bullock}, J.~S., {Zentner}, A.~R., {Kravtsov}, A.~V., \&
  {Moustakas}, L.~A. 2008,
  \href{http://dx.doi.org/10.1086/591958}{\JournalTitle{\apj}, 688, 254}

\bibitem[{{Kordopatis} {et~al.}(2015){Kordopatis}, {Binney}, {Gilmore}, {Wyse},
  {Belokurov}, {McMillan}, {Hatfield}, {Grebel}, {Steinmetz}, {Navarro},
  {Seabroke}, {Minchev}, {Chiappini}, {Bienaym{\'e}}, {Bland-Hawthorn},
  {Freeman}, {Gibson}, {Helmi}, {Munari}, {Parker}, {Reid}, {Siebert},
  {Siviero}, \& {Zwitter}}]{2015MNRAS.447.3526K}
{Kordopatis}, G., {Binney}, J., {Gilmore}, G., {et~al.} 2015,
  \href{http://dx.doi.org/10.1093/mnras/stu2726}{\JournalTitle{\mnras}, 447,
  3526}

\bibitem[{{Lacey}(1984)}]{1984MNRAS.208..687L}
{Lacey}, C.~G. 1984,
  \href{http://dx.doi.org/10.1093/mnras/208.4.687}{\JournalTitle{\mnras}, 208,
  687}

\bibitem[{{Lindegren} {et~al.}(2018){Lindegren}, {Hern{\'a}ndez}, {Bombrun},
  {Klioner}, {Bastian}, {Ramos-Lerate}, {de Torres}, {Steidelm{\"u}ller},
  {Stephenson}, {Hobbs}, {Lammers}, {Biermann}, {Geyer}, {Hilger}, {Michalik},
  {Stampa}, {McMillan}, {Casta{\~n}eda}, {Clotet}, {Comoretto}, {Davidson},
  {Fabricius}, {Gracia}, {Hambly}, {Hutton}, {Mora}, {Portell}, {van Leeuwen},
  {Abbas}, {Abreu}, {Altmann}, {Andrei}, {Anglada}, {Balaguer-N{\'u}{\~n}ez},
  {Barache}, {Becciani}, {Bertone}, {Bianchi}, {Bouquillon}, {Bourda},
  {Br{\"u}semeister}, {Bucciarelli}, {Busonero}, {Buzzi}, {Cancelliere},
  {Carlucci}, {Charlot}, {Cheek}, {Crosta}, {Crowley}, {de Bruijne}, {de
  Felice}, {Drimmel}, {Esquej}, {Fienga}, {Fraile}, {Gai}, {Garralda},
  {Gonz{\'a}lez-Vidal}, {Guerra}, {Hauser}, {Hofmann}, {Holl}, {Jordan},
  {Lattanzi}, {Lenhardt}, {Liao}, {Licata}, {Lister}, {L{\"o}ffler},
  {Marchant}, {Martin-Fleitas}, {Messineo}, {Mignard}, {Morbidelli}, {Poggio},
  {Riva}, {Rowell}, {Salguero}, {Sarasso}, {Sciacca}, {Siddiqui}, {Smart},
  {Spagna}, {Steele}, {Taris}, {Torra}, {van Elteren}, {van Reeven}, \&
  {Vecchiato}}]{2018A&A...616A...2L}
{Lindegren}, L., {Hern{\'a}ndez}, J., {Bombrun}, A., {et~al.} 2018,
  \href{http://dx.doi.org/10.1051/0004-6361/201832727}{\JournalTitle{\aap},
  616, A2}

\bibitem[{{Mackereth} {et~al.}(2017){Mackereth}, {Bovy}, {Schiavon},
  {Zasowski}, {Cunha}, {Frinchaboy}, {Garc{\'\i}a Perez}, {Hayden}, {Holtzman},
  {Majewski}, {M{\'e}sz{\'a}ros}, {Nidever}, {Pinsonneault}, \&
  {Shetrone}}]{2017MNRAS.471.3057M}
{Mackereth}, J.~T., {Bovy}, J., {Schiavon}, R.~P., {et~al.} 2017,
  \href{http://dx.doi.org/10.1093/mnras/stx1774}{\JournalTitle{\mnras}, 471,
  3057}

\bibitem[{{Mackereth} {et~al.}(2019){Mackereth}, {Bovy}, {Leung}, {Schiavon},
  {Trick}, {Chaplin}, {Cunha}, {Feuillet}, {Majewski}, {Martig}, {Miglio},
  {Nidever}, {Pinsonneault}, {Aguirre}, {Sobeck}, {Tayar}, \&
  {Zasowski}}]{2019MNRAS.489..176M}
{Mackereth}, J.~T., {Bovy}, J., {Leung}, H.~W., {et~al.} 2019,
  \href{http://dx.doi.org/10.1093/mnras/stz1521}{\JournalTitle{\mnras}, 489,
  176}

\bibitem[{{Majewski} {et~al.}(2017){Majewski}, {Schiavon}, {Frinchaboy},
  {Allende Prieto}, {Barkhouser}, {Bizyaev}, {Blank}, {Brunner}, {Burton},
  {Carrera}, {Chojnowski}, {Cunha}, {Epstein}, {Fitzgerald}, {Garc{\'\i}a
  P{\'e}rez}, {Hearty}, {Henderson}, {Holtzman}, {Johnson}, {Lam}, {Lawler},
  {Maseman}, {M{\'e}sz{\'a}ros}, {Nelson}, {Nguyen}, {Nidever}, {Pinsonneault},
  {Shetrone}, {Smee}, {Smith}, {Stolberg}, {Skrutskie}, {Walker}, {Wilson},
  {Zasowski}, {Anders}, {Basu}, {Beland}, {Blanton}, {Bovy}, {Brownstein},
  {Carlberg}, {Chaplin}, {Chiappini}, {Eisenstein}, {Elsworth}, {Feuillet},
  {Fleming}, {Galbraith-Frew}, {Garc{\'\i}a}, {Garc{\'\i}a-Hern{\'a}ndez},
  {Gillespie}, {Girardi}, {Gunn}, {Hasselquist}, {Hayden}, {Hekker}, {Ivans},
  {Kinemuchi}, {Klaene}, {Mahadevan}, {Mathur}, {Mosser}, {Muna}, {Munn},
  {Nichol}, {O'Connell}, {Parejko}, {Robin}, {Rocha-Pinto}, {Schultheis},
  {Serenelli}, {Shane}, {Silva Aguirre}, {Sobeck}, {Thompson}, {Troup},
  {Weinberg}, \& {Zamora}}]{2017AJ....154...94M}
{Majewski}, S.~R., {Schiavon}, R.~P., {Frinchaboy}, P.~M., {et~al.} 2017,
  \href{http://dx.doi.org/10.3847/1538-3881/aa784d}{\JournalTitle{\aj}, 154,
  94}

\bibitem[{{Marigo} {et~al.}(2017){Marigo}, {Girardi}, {Bressan}, {Rosenfield},
  {Aringer}, {Chen}, {Dussin}, {Nanni}, {Pastorelli}, {Rodrigues}, {Trabucchi},
  {Bladh}, {Dalcanton}, {Groenewegen}, {Montalb{\'a}n}, \&
  {Wood}}]{2017ApJ...835...77M}
{Marigo}, P., {Girardi}, L., {Bressan}, A., {et~al.} 2017,
  \href{http://dx.doi.org/10.3847/1538-4357/835/1/77}{\JournalTitle{\apj}, 835,
  77}

\bibitem[{{Martinez-Medina} {et~al.}(2015){Martinez-Medina}, {Pichardo},
  {P{\'e}rez-Villegas}, \& {Moreno}}]{2015ApJ...802..109M}
{Martinez-Medina}, L.~A., {Pichardo}, B., {P{\'e}rez-Villegas}, A., \&
  {Moreno}, E. 2015,
  \href{http://dx.doi.org/10.1088/0004-637X/802/2/109}{\JournalTitle{\apj},
  802, 109}

\bibitem[{{Minchev}(2016)}]{2016AN....337..703M}
{Minchev}, I. 2016,
  \href{http://dx.doi.org/10.1002/asna.201612366}{\JournalTitle{Astronomische
  Nachrichten}, 337, 703}

\bibitem[{{Minchev} {et~al.}(2012){Minchev}, {Famaey}, {Quillen}, {Dehnen},
  {Martig}, \& {Siebert}}]{2012A&A...548A.127M}
{Minchev}, I., {Famaey}, B., {Quillen}, A.~C., {et~al.} 2012,
  \href{http://dx.doi.org/10.1051/0004-6361/201219714}{\JournalTitle{\aap},
  548, A127}

\bibitem[{{Minchev} {et~al.}(2018){Minchev}, {Anders}, {Recio-Blanco},
  {Chiappini}, {de Laverny}, {Queiroz}, {Steinmetz}, {Adibekyan}, {Carrillo},
  {Cescutti}, {Guiglion}, {Hayden}, {de Jong}, {Kordopatis}, {Majewski},
  {Martig}, \& {Santiago}}]{2018MNRAS.481.1645M}
{Minchev}, I., {Anders}, F., {Recio-Blanco}, A., {et~al.} 2018,
  \href{http://dx.doi.org/10.1093/mnras/sty2033}{\JournalTitle{\mnras}, 481,
  1645}

\bibitem[{{Nordstr{\"o}m} {et~al.}(2004){Nordstr{\"o}m}, {Mayor}, {Andersen},
  {Holmberg}, {Pont}, {J{\o}rgensen}, {Olsen}, {Udry}, \&
  {Mowlavi}}]{2004A&A...418..989N}
{Nordstr{\"o}m}, B., {Mayor}, M., {Andersen}, J., {et~al.} 2004,
  \href{http://dx.doi.org/10.1051/0004-6361:20035959}{\JournalTitle{\aap}, 418,
  989}

\bibitem[{{Pinsonneault} {et~al.}(2018){Pinsonneault}, {Elsworth}, {Tayar},
  {Serenelli}, {Stello}, {Zinn}, {Mathur}, {Garc{\'\i}a}, {Johnson}, {Hekker},
  {Huber}, {Kallinger}, {M{\'e}sz{\'a}ros}, {Mosser}, {Stassun}, {Girardi},
  {Rodrigues}, {Silva Aguirre}, {An}, {Basu}, {Chaplin}, {Corsaro}, {Cunha},
  {Garc{\'\i}a-Hern{\'a}ndez}, {Holtzman}, {J{\"o}nsson}, {Shetrone}, {Smith},
  {Sobeck}, {Stringfellow}, {Zamora}, {Beers}, {Fern{\'a}ndez-Trincado},
  {Frinchaboy}, {Hearty}, \& {Nitschelm}}]{2018ApJS..239...32P}
{Pinsonneault}, M.~H., {Elsworth}, Y.~P., {Tayar}, J., {et~al.} 2018,
  \href{http://dx.doi.org/10.3847/1538-4365/aaebfd}{\JournalTitle{\apjs}, 239,
  32}

\bibitem[{{Piskunov} \& {Valenti}(2017)}]{2017A&A...597A..16P}
{Piskunov}, N., \& {Valenti}, J.~A. 2017,
  \href{http://dx.doi.org/10.1051/0004-6361/201629124}{\JournalTitle{\aap},
  597, A16}

\bibitem[{{Quillen} \& {Garnett}(2001)}]{2001ASPC..230...87Q}
{Quillen}, A.~C., \& {Garnett}, D.~R. 2001, in Astronomical Society of the
  Pacific Conference Series, Vol. 230, Galaxy Disks and Disk Galaxies, ed.
  J.~G. {Funes} \& E.~M. {Corsini} (Astronomical Society of the Pacific), 87

\bibitem[{{Reid}(1993)}]{1993ARA&A..31..345R}
{Reid}, M.~J. 1993,
  \href{http://dx.doi.org/10.1146/annurev.aa.31.090193.002021}{\JournalTitle{\araa},
  31, 345}

\bibitem[{{Reid} \& {Brunthaler}(2004)}]{2004ApJ...616..872R}
{Reid}, M.~J., \& {Brunthaler}, A. 2004,
  \href{http://dx.doi.org/10.1086/424960}{\JournalTitle{\apj}, 616, 872}

\bibitem[{{Robin} {et~al.}(2003){Robin}, {Reyl{\'e}}, {Derri{\`e}re}, \&
  {Picaud}}]{2003A&A...409..523R}
{Robin}, A.~C., {Reyl{\'e}}, C., {Derri{\`e}re}, S., \& {Picaud}, S. 2003,
  \href{http://dx.doi.org/10.1051/0004-6361:20031117}{\JournalTitle{\aap}, 409,
  523}

\bibitem[{{Ro{\v{s}}kar} {et~al.}(2010){Ro{\v{s}}kar}, {Debattista}, {Brooks},
  {Quinn}, {Brook}, {Governato}, {Dalcanton}, \&
  {Wadsley}}]{2010MNRAS.408..783R}
{Ro{\v{s}}kar}, R., {Debattista}, V.~P., {Brooks}, A.~M., {et~al.} 2010,
  \href{http://dx.doi.org/10.1111/j.1365-2966.2010.17178.x}{\JournalTitle{\mnras},
  408, 783}

\bibitem[{{Ro{\v{s}}kar} {et~al.}(2008){Ro{\v{s}}kar}, {Debattista}, {Quinn},
  {Stinson}, \& {Wadsley}}]{2008ApJ...684L..79R}
{Ro{\v{s}}kar}, R., {Debattista}, V.~P., {Quinn}, T.~R., {Stinson}, G.~S., \&
  {Wadsley}, J. 2008,
  \href{http://dx.doi.org/10.1086/592231}{\JournalTitle{\apjl}, 684, L79}

\bibitem[{{Saha} {et~al.}(2010){Saha}, {Tseng}, \&
  {Taam}}]{2010ApJ...721.1878S}
{Saha}, K., {Tseng}, Y.-H., \& {Taam}, R.~E. 2010,
  \href{http://dx.doi.org/10.1088/0004-637X/721/2/1878}{\JournalTitle{\apj},
  721, 1878}

\bibitem[{{Sanders} \& {Binney}(2015)}]{2015MNRAS.449.3479S}
{Sanders}, J.~L., \& {Binney}, J. 2015,
  \href{http://dx.doi.org/10.1093/mnras/stv578}{\JournalTitle{\mnras}, 449,
  3479}

\bibitem[{{Sanders} \& {Das}(2018)}]{2018MNRAS.481.4093S}
{Sanders}, J.~L., \& {Das}, P. 2018,
  \href{http://dx.doi.org/10.1093/mnras/sty2490}{\JournalTitle{\mnras}, 481,
  4093}

\bibitem[{{Sch{\"o}nrich} \& {Binney}(2009)}]{2009MNRAS.396..203S}
{Sch{\"o}nrich}, R., \& {Binney}, J. 2009,
  \href{http://dx.doi.org/10.1111/j.1365-2966.2009.14750.x}{\JournalTitle{\mnras},
  396, 203}

\bibitem[{{Sellwood}(2008)}]{2008ASPC..396..341S}
{Sellwood}, J.~A. 2008, in Astronomical Society of the Pacific Conference
  Series, Vol. 396, Formation and Evolution of Galaxy Disks, ed. J.~G. {Funes}
  \& E.~M. {Corsini}, 341

\bibitem[{{Sellwood}(2013)}]{2013ApJ...769L..24S}
{Sellwood}, J.~A. 2013,
  \href{http://dx.doi.org/10.1088/2041-8205/769/2/L24}{\JournalTitle{\apjl},
  769, L24}

\bibitem[{{Sellwood} \& {Binney}(2002)}]{2002MNRAS.336..785S}
{Sellwood}, J.~A., \& {Binney}, J.~J. 2002,
  \href{http://dx.doi.org/10.1046/j.1365-8711.2002.05806.x}{\JournalTitle{\mnras},
  336, 785}

\bibitem[{{Sellwood} \& {Carlberg}(1984)}]{1984ApJ...282...61S}
{Sellwood}, J.~A., \& {Carlberg}, R.~G. 1984,
  \href{http://dx.doi.org/10.1086/162176}{\JournalTitle{\apj}, 282, 61}

\bibitem[{{Sharma} {et~al.}(2012){Sharma}, {Steinmetz}, \&
  {Bland-Hawthorn}}]{2012ApJ...750..107S}
{Sharma}, S., {Steinmetz}, M., \& {Bland-Hawthorn}, J. 2012,
  \href{http://dx.doi.org/10.1088/0004-637X/750/2/107}{\JournalTitle{\apj},
  750, 107}

\bibitem[{{Sharma} {et~al.}(2016){Sharma}, {Stello}, {Bland-Hawthorn}, {Huber},
  \& {Bedding}}]{2016ApJ...822...15S}
{Sharma}, S., {Stello}, D., {Bland-Hawthorn}, J., {Huber}, D., \& {Bedding},
  T.~R. 2016,
  \href{http://dx.doi.org/10.3847/0004-637X/822/1/15}{\JournalTitle{\apj}, 822,
  15}

\bibitem[{{Sharma} {et~al.}(2017){Sharma}, {Stello}, {Huber}, {Bland
  -Hawthorn}, \& {Bedding}}]{2017ApJ...835..163S}
{Sharma}, S., {Stello}, D., {Huber}, D., {Bland -Hawthorn}, J., \& {Bedding},
  T.~R. 2017,
  \href{http://dx.doi.org/10.3847/1538-4357/835/2/163}{\JournalTitle{\apj},
  835, 163}

\bibitem[{{Sharma} {et~al.}(2014){Sharma}, {Bland-Hawthorn}, {Binney},
  {Freeman}, {Steinmetz}, {Boeche}, {Bienaym{\'e}}, {Gibson}, {Gilmore},
  {Grebel}, {Helmi}, {Kordopatis}, {Munari}, {Navarro}, {Parker}, {Reid},
  {Seabroke}, {Siebert}, {Watson}, {Williams}, {Wyse}, \&
  {Zwitter}}]{2014ApJ...793...51S}
{Sharma}, S., {Bland-Hawthorn}, J., {Binney}, J., {et~al.} 2014,
  \href{http://dx.doi.org/10.1088/0004-637X/793/1/51}{\JournalTitle{\apj}, 793,
  51}

\bibitem[{{Sharma} {et~al.}(2018){Sharma}, {Stello}, {Buder}, {Kos},
  {Bland-Hawthorn}, {Asplund}, {Duong}, {Lin}, {Lind}, {Ness}, {Huber},
  {Zwitter}, {Traven}, {Hon}, {Kafle}, {Khanna}, {Saddon}, {Anguiano}, {Casey},
  {Freeman}, {Martell}, {De Silva}, {Simpson}, {Wittenmyer}, \&
  {Zucker}}]{2018MNRAS.473.2004S}
{Sharma}, S., {Stello}, D., {Buder}, S., {et~al.} 2018,
  \href{http://dx.doi.org/10.1093/mnras/stx2582}{\JournalTitle{\mnras}, 473,
  2004}

\bibitem[{{Sharma} {et~al.}(2019){Sharma}, {Stello}, {Bland-Hawthorn},
  {Hayden}, {Zinn}, {Kallinger}, {Hon}, {Asplund}, {Buder}, {De Silva},
  {D'Orazi}, {Freeman}, {Kos}, {Lewis}, {Lin}, {Lind}, {Martell}, {Simpson},
  {Wittenmyer}, {Zucker}, {Zwitter}, {Bedding}, {Chen}, {Cotar}, {Esdaile},
  {Horner}, {Huber}, {Kafle}, {Khanna}, {Li}, {Ting}, {Nataf}, {Nordlander},
  {Saadon}, {Traven}, {Wright}, \& {Wyse}}]{2019MNRAS.490.5335S}
{Sharma}, S., {Stello}, D., {Bland-Hawthorn}, J., {et~al.} 2019,
  \href{http://dx.doi.org/10.1093/mnras/stz2861}{\JournalTitle{\mnras}, 490,
  5335}

\bibitem[{{Shiidsuka} \& {Ida}(1999)}]{1999MNRAS.307..737S}
{Shiidsuka}, K., \& {Ida}, S. 1999,
  \href{http://dx.doi.org/10.1046/j.1365-8711.1999.02598.x}{\JournalTitle{\mnras},
  307, 737}

\bibitem[{{Silva Aguirre} {et~al.}(2018){Silva Aguirre}, {Bojsen-Hansen},
  {Slumstrup}, {Casagrande}, {Kawata}, {Ciuc{\v{a}}}, {Hand berg}, {Lund},
  {Mosumgaard}, {Huber}, {Johnson}, {Pinsonneault}, {Serenelli}, {Stello},
  {Tayar}, {Bird}, {Cassisi}, {Hon}, {Martig}, {Nissen}, {Rix},
  {Sch{\"o}nrich}, {Sahlholdt}, {Trick}, \& {Yu}}]{2018MNRAS.475.5487S}
{Silva Aguirre}, V., {Bojsen-Hansen}, M., {Slumstrup}, D., {et~al.} 2018,
  \href{http://dx.doi.org/10.1093/mnras/sty150}{\JournalTitle{\mnras}, 475,
  5487}

\bibitem[{{Solway} {et~al.}(2012){Solway}, {Sellwood}, \&
  {Sch{\"o}nrich}}]{2012MNRAS.422.1363S}
{Solway}, M., {Sellwood}, J.~A., \& {Sch{\"o}nrich}, R. 2012,
  \href{http://dx.doi.org/10.1111/j.1365-2966.2012.20712.x}{\JournalTitle{\mnras},
  422, 1363}

\bibitem[{{Spitzer} \& {Schwarzschild}(1951)}]{1951ApJ...114..385S}
{Spitzer}, Lyman, J., \& {Schwarzschild}, M. 1951,
  \href{http://dx.doi.org/10.1086/145478}{\JournalTitle{\apj}, 114, 385}

\bibitem[{{Stello} {et~al.}(2015){Stello}, {Huber}, {Sharma}, {Johnson},
  {Lund}, {Handberg}, {Buzasi}, {Silva Aguirre}, {Chaplin}, {Miglio},
  {Pinsonneault}, {Basu}, {Bedding}, {Bland-Hawthorn}, {Casagrande}, {Davies},
  {Elsworth}, {Garcia}, {Mathur}, {Di Mauro}, {Mosser}, {Schneider},
  {Serenelli}, \& {Valentini}}]{2015ApJ...809L...3S}
{Stello}, D., {Huber}, D., {Sharma}, S., {et~al.} 2015,
  \href{http://dx.doi.org/10.1088/2041-8205/809/1/L3}{\JournalTitle{\apjl},
  809, L3}

\bibitem[{{Ting} \& {Rix}(2019)}]{2019ApJ...878...21T}
{Ting}, Y.-S., \& {Rix}, H.-W. 2019,
  \href{http://dx.doi.org/10.3847/1538-4357/ab1ea5}{\JournalTitle{\apj}, 878,
  21}

\bibitem[{{Valenti} \& {Piskunov}(1996)}]{1996A&AS..118..595V}
{Valenti}, J.~A., \& {Piskunov}, N. 1996, \JournalTitle{\aaps}, 118, 595

\bibitem[{{van der Kruit}(1988)}]{1988A&A...192..117V}
{van der Kruit}, P.~C. 1988, \JournalTitle{\aap}, 192, 117

\bibitem[{{Velazquez} \& {White}(1999)}]{1999MNRAS.304..254V}
{Velazquez}, H., \& {White}, S. D.~M. 1999,
  \href{http://dx.doi.org/10.1046/j.1365-8711.1999.02354.x}{\JournalTitle{\mnras},
  304, 254}

\bibitem[{{Vera-Ciro} {et~al.}(2014){Vera-Ciro}, {D'Onghia}, {Navarro}, \&
  {Abadi}}]{2014ApJ...794..173V}
{Vera-Ciro}, C., {D'Onghia}, E., {Navarro}, J., \& {Abadi}, M. 2014,
  \href{http://dx.doi.org/10.1088/0004-637X/794/2/173}{\JournalTitle{\apj},
  794, 173}

\bibitem[{{Villalobos} \& {Helmi}(2008)}]{2008MNRAS.391.1806V}
{Villalobos}, {\'A}., \& {Helmi}, A. 2008,
  \href{http://dx.doi.org/10.1111/j.1365-2966.2008.13979.x}{\JournalTitle{\mnras},
  391, 1806}

\bibitem[{{Wisnioski} {et~al.}(2015){Wisnioski}, {F{\"o}rster Schreiber},
  {Wuyts}, {Wuyts}, {Bandara}, {Wilman}, {Genzel}, {Bender}, {Davies},
  {Fossati}, {Lang}, {Mendel}, {Beifiori}, {Brammer}, {Chan}, {Fabricius},
  {Fudamoto}, {Kulkarni}, {Kurk}, {Lutz}, {Nelson}, {Momcheva}, {Rosario},
  {Saglia}, {Seitz}, {Tacconi}, \& {van Dokkum}}]{2015ApJ...799..209W}
{Wisnioski}, E., {F{\"o}rster Schreiber}, N.~M., {Wuyts}, S., {et~al.} 2015,
  \href{http://dx.doi.org/10.1088/0004-637X/799/2/209}{\JournalTitle{\apj},
  799, 209}

\bibitem[{{Wu} {et~al.}(2019){Wu}, {Xiang}, {Zhao}, {Bi}, {Liu}, {Shi},
  {Huang}, {Yuan}, {Wang}, {Chen}, {Huo}, {Ren}, {Tian}, {Liu}, {Zhang}, {Li},
  \& {Zhang}}]{2019MNRAS.484.5315W}
{Wu}, Y., {Xiang}, M., {Zhao}, G., {et~al.} 2019,
  \href{http://dx.doi.org/10.1093/mnras/stz256}{\JournalTitle{\mnras}, 484,
  5315}

\bibitem[{{Xiang} {et~al.}(2017{\natexlab{a}}){Xiang}, {Liu}, {Shi}, {Yuan},
  {Huang}, {Chen}, {Wang}, {Tian}, {Wu}, {Yang}, {Zhang}, {Huo}, \&
  {Ren}}]{2017ApJS..232....2X}
{Xiang}, M., {Liu}, X., {Shi}, J., {et~al.} 2017{\natexlab{a}},
  \href{http://dx.doi.org/10.3847/1538-4365/aa80e4}{\JournalTitle{\apjs}, 232,
  2}

\bibitem[{{Xiang} {et~al.}(2017{\natexlab{b}}){Xiang}, {Liu}, {Yuan}, {Huo},
  {Huang}, {Wang}, {Chen}, {Ren}, {Zhang}, {Tian}, {Yang}, {Shi}, {Zhao}, {Li},
  {Zhao}, {Cui}, {Li}, {Hou}, {Zhang}, {Zhang}, {Wang}, {Wu}, {Cao}, {Yan},
  {Yan}, {Luo}, {Zhang}, {Bai}, {Yuan}, {Dong}, {Lei}, \&
  {Li}}]{2017MNRAS.467.1890X}
{Xiang}, M.~S., {Liu}, X.~W., {Yuan}, H.~B., {et~al.} 2017{\natexlab{b}},
  \href{http://dx.doi.org/10.1093/mnras/stx129}{\JournalTitle{\mnras}, 467,
  1890}

\bibitem[{{Zhao} {et~al.}(2012){Zhao}, {Zhao}, {Chu}, {Jing}, \&
  {Deng}}]{2012RAA....12..723Z}
{Zhao}, G., {Zhao}, Y.-H., {Chu}, Y.-Q., {Jing}, Y.-P., \& {Deng}, L.-C. 2012,
  \href{http://dx.doi.org/10.1088/1674-4527/12/7/002}{\JournalTitle{Research in
  Astronomy and Astrophysics}, 12, 723}

\end{thebibliography}

\end{document}